\documentclass[fleqn,usenatbib]{mnras}

\usepackage{newtxtext,newtxmath}

\usepackage[T1]{fontenc}
\usepackage{ae,aecompl}

\usepackage{graphicx}	
\usepackage{multicol}

\newcommand{\sgBe}{B[\,e\,] supergiant}
\newcommand{\sgBeshorthand}{B[\,e\,]SG}
\newcommand{\kms}{$\rm km\,s^{-1}$}

\title[GG Carinae: variability at the orbital period]{GG Carinae: orbital parameters and accretion indicators from phase-resolved spectroscopy and photometry}

\author[A. J. D. Porter et al.]{
Augustus Porter$^{1}$\thanks{E-mail: augustus.porter@physics.ox.ac.uk}, 
David Grant$^{1}$,
Katherine Blundell$^{1}$,
and Steven Lee$^{2,3}$
\\
$^{1}$Department of Physics, University of Oxford, Denys Wilkinson Building, Oxford, OX1 3RH, United Kingdom\\
$^{2}$Anglo-Australian Telescope, Coonabarabran NSW 2357, Australia\\
$^3$Research School of Astronomy and Astrophysics, Australian National University, Canberra, ACT 2611
}

\date{Accepted 2020 November 26. Received 2020 November 19; in original form 2020 October 05}

\pubyear{2020}

\hypersetup{final}
\begin{document}
\label{firstpage}
\pagerange{\pageref{firstpage}--\pageref{lastpage}}
\maketitle

\begin{abstract}
    \textcolor{black}{B[\,e\,] supergiants are a rare and unusual class of massive and luminous stars, characterised by opaque circumstellar envelopes. GG Carinae is a binary whose primary component is a B[\,e\,] supergiant and whose variability has remained unsatisfactorily explained. Using photometric data from ASAS, OMC, and ASAS-SN, and spectroscopic data from the Global Jet Watch and FEROS to study visible emission lines, we focus on the variability of the system at its $\sim$31-day orbital period and constrain the stellar parameters of the primary. There is one photometric minimum per orbital period and, in the emission line spectroscopy, we find a correlation between the amplitude of radial velocity variations and the initial energy of the line species. The spectral behaviour is consistent with the emission lines forming in the primary's wind, with the variable amplitudes between line species being caused by the less energetic lines forming at larger radii on average. By modelling the atmosphere of the primary, we are able to model the radial velocity variations of the wind lines in order to constrain the orbit of the binary. We find that the binary is even more eccentric than previously believed ($e=0.5\pm0.03$). Using this orbital solution, the system is brightest at periastron and dimmest at apastron, and the shape of the photometric variations at the orbital period can be well described by the variable accretion by the secondary of the primary's wind. We suggest that the evolutionary history of GG Carinae may need to be reevaluated in a binary context.}
\end{abstract}




\section{Introduction}
\label{sec:introduction}
\sgBe s (\sgBeshorthand s) are a class of rare stars which are not predicted by stellar evolution models. They are characterized by hybrid spectra of hot stars with infrared excess; strong emission in Hydrogen Balmer and Helium lines; strong permitted and forbidden emission lines from a number of elements; and wide absorption lines in the ultraviolet (UV) spectrum. These features point towards a complex circumstellar environment \citep{Zickgraf1985, Zickgraf1986, Kraus2019ASupergiants}. Currently there are only $\sim$33 confirmed \sgBeshorthand s discovered, and $\sim$25 further candidates \citep{Kraus2014DISCOVERY31, Levato2014NewCloud, Kraus2009, Kraus, Kraus2019ASupergiants}. Their formation channels and the origin of the B[\,e\,] phenomenon are unclear, with some studies ascribing the phenomena to binarity \citep{Podsiadlowski2006, Miroshnichenko2007, Wang2012AMBER/VLTI300} and others to non-radial pulsations \citep{Kraus2016a}. The opaque circumstellar envelopes of \sgBeshorthand s preclude the observation of photospheric absorption lines and therefore the determination of the stars' surface conditions (e.g. \citealt{Kraus2009}). The exact mechanism of mass-loss in these stars remains unknown.  \\

Akin to classical Be stars, the standard picture of \sgBeshorthand s are that they are expected to be rapid rotators which lead to equatorial outflows \citep{Zickgraf1986}; however their lack of photospheric lines make it difficult to confirm this for the majority of the population sample. For the few \sgBeshorthand s for which rotation rates have been measured, they have been found to rotate at a high fraction of the critical break-up velocity \citep{Gummersbach1995BeClouds, Zickgraf1999TheStars, Zickgraf2006BeClouds, Kraus2008}. However, doubt has been cast on the measurement of these rotation velocities by \cite{Kraus2016}, who found the He\,I absorption used to determine the rotation velocity of one specimen was polluted by the wind absorption and by variability associated with pulsations. They conclude that this casts doubt on the rotation velocities of the entire sample.\\

\textcolor{black}{\sgBeshorthand s are usually not significantly photometrically variable, except for binary induced variability \citep{Zickgraf1986, Lamers1998}. \cite{Krtickova2018AnEnvelopes} suggests that any long-term broad-band photometric variability in \sgBeshorthand s is most likely due to changes in the stars' envelopes.} \\

GG Carinae (GG Car, also known as HD 94878 and CPD-59 2855), is an enigmatic Galactic \sgBeshorthand\ binary which has been studied for over a century due to its peculiar spectroscopic and photometric properties \citep{Pickering1896HarvardSpectra., Kruytbosch1930Variability2855, Greenstein1938FourCarinae}. \cite{Lamers1998} classified GG Car as a \sgBeshorthand, building on the work of \cite{Mcgregor1988ATOMICSTARS} and \cite{LOPES1992AStars}, noting their observation of the B[\,e\,] phenomenon in this object: its high luminosity, indications of mass-loss through P Cygni line profiles, and its hosting of a hybrid spectrum of narrow emission lines and broad absorption features. \cite{Marchiano2012} refined the luminosity and the temperature of the primary, cementing GG Car's status as a \sgBeshorthand. CO emission present in the system suggests that the primary is in an early, pre-red supergiant, evolved stage of its post-main sequence evolution \citep{Kraus2009, Kraus2013, Oksala2013ProbingTransition}. However, this determination depends on assumptions of the rotation rate of the primary, which cannot be directly measured. \\

The binarity of GG Car has long been hinted at by photometry \citep{Kruytbosch1930Variability2855, Gosset1984}, and has been confirmed spectroscopically by the periodic radial velocities of its emission lines \citep{Hernandez1981, Gosset1985}. It had been debated whether the system has a photometric period of $\sim$31 or $\sim$62 days \citep{Kruytbosch1930Variability2855, Greenstein1938FourCarinae, Gosset1984, VanLeeuwen1998HipparcosVariables}, whereas the spectroscopic variability indubitably displays $\sim$31-day periodicity \citep{Hernandez1981, Gosset1985, Marchiano2012}. \textcolor{black}{However, despite these studies being able to determine the binary's orbital period, an accurate determination of GG Car's binary orbital solution has remained elusive due to the lack of photospheric absorption lines in the system, typical of \sgBeshorthand s. The confusion as to the photometric period arises from long- and short-period variability in the system, which causes successive orbits' lightcurves to be non-identical \citep{Gosset1984, VanLeeuwen1998HipparcosVariables, Krtickova2018AnEnvelopes}.  This variability makes GG Car unusual for a \sgBeshorthand.}\\

The class of the secondary component of GG Car is undetermined. The system was observed in the X-ray band by the SWIFT mission, and the data are available in the 1SXPS catalogue \citep{Evans20131SXPS:SPECTRA}. There was a ``poor'' detection of X-rays from the system, not significantly above the background level. This indicates that GG Car is not a significant source of X-ray radiation, and it is unlikely the system is an X-ray binary or a highly energetic colliding-wind binary such as $\eta$ Carinae \citep{Pittard1998TheCarinae}. \\

\cite{Kraus2013} discovered infrared emission originating from a thin CO circumbinary ring orbiting at 80\,${\rm km\, s^{-1}}$ projected onto the line of sight, and suggested that the circumbinary disk was formed during a classical Be phase of the primary before the star exhibited the B[\,e\,] phenomena. \cite{Maravelias2018} studied GG Car's forbidden optical emission lines and found [Ca\,II] and [O\,I] emission lines originating from the same region as the CO emission, and another [O\,I] ring orbiting with a projected velocity of $\sim$30\,\kms. They determined the systemic velocity of the system to be $-22$\,km\,s$^{-1}$ using these circumbinary lines. \cite{Pereyra2009} found that GG Car hosts a rotating disk emitting H-alpha, and by consideration of the time-varying spectropolarimetry concluded that it originates at least in part in the system's circumbinary disk. \\

\textcolor{black}{The standard picture of \sgBeshorthand s is that they host two wind components: a hot, fast polar wind, and a slow, cool equatorial wind \citep{Zickgraf1985, Zickgraf1986}. \cite{Oudmaijer1998TheStars} extended this, finding that at intermediate latitudes the wind had properties between these two extremes, indicating a less clear cut picture but that complex wind absorption components could be observed. \cite{Gosset1985} noted that there were multiple blue shifted absorption components in the H\,I Balmer and He\,I emission lines, with the number of components depending on orbital phase. Additionally, they determined that the orbital solution of the binary is dependent on the line species studied which, without proper physical elucidation, undermines confidence in the solution.}\\

\textcolor{black}{\cite{Marchiano2012} interpreted the He\,I absorption components in GG Car as photospheric absorption from both of the binary components, and from this determined a measurement of GG Car's orbital parameters. This interpretation of the He\,I absorption would be unusual for \sgBeshorthand s, and that interpretation is highly suspect when the He\,I line profiles include both emission and absorption from the complex wind of the primary. With their interpretation, \cite{Marchiano2012} inferred a systemic velocity of GG Car of $V_r = -162$\,\kms\ from the He\,I absorption lines. This systemic velocity is incompatible with the $-22$\,\kms\ determined by \cite{Maravelias2018}. Additionally, \cite{Hanes2018StellarNebula} measured that the average systemic velocity $\langle V_r \rangle = -7.14 \pm 13.10$\,\kms\ for the stars composing the Carina nebula. This indicates that the He\,I absorption of GG Car measured by \cite{Marchiano2012} were systematically more blueshifted than the systemic velocity, and therefore these absorptions are in fact due to absorption in the complex wind of the system. This casts doubt on the accuracy of the orbital solution of the binary that they measured; therefore, in this study we aim to uncover the true binary orbital solution in the system using alternate methods.} \\

With its luminosity, high mass, and complex circumstellar environment, GG Car appears to be the prototypical \sgBeshorthand\ system; however, its variability on both orbital and shorter timescales have shown that it is an unusual system. In this study, we investigate the variability of GG Car at its $\sim$31-day orbital period, using both photometric and spectroscopic data. In Section \ref{sec:observations}, we introduce the V-band photometry and the Global Jet Watch and FEROS spectroscopy of GG Car. In Section \ref{sec:gaia_luminosity}, we determine the luminosity of the primary using the \textit{Gaia} parallax of the system and estimate the primary's mass. In Section \ref{sec:variability}, we study the variability of the system's photometry and emission lines' radial velocities at the $\sim$31 day orbital period. \textcolor{black}{ In Section \ref{sec:determining_orbital_solution}, we constrain the orbital solution of the binary by properly calculating the time-dependent emissivity of GG Car's emission lines in its atmosphere and using them to fit the RV variations of the emission lines. We find a new orbital ephemeris of}
\begin{equation}
\label{eq:ephemeris}
    p = \frac{T - 2452069.36}{31.011}
\end{equation}
\noindent \textcolor{black}{where $p$ is the phase and $T$ is the time in JD. Therefore, phase $0/1$ corresponds with periastron of the binary, and phase $0.5$ corresponds to apastron. All phases presented in this study are calculated using Equation \ref{eq:ephemeris}.} In Section \ref{sec:discussion} we constrain the mass of the secondary component, and then discuss some models which may cause the photometric variability seen in GG Car, and in Section \ref{sec:conclusions} we present our conclusions and summarise the properties of GG Car that we find in this study. \\

This work is the first in a series of studies of GG Car which will investigate short-period, 1.583-day variability in the system, and the atomic circumbinary emission in the visible spectrum of GG Car. \\

\section{Observations}
\label{sec:observations}
\subsection{V-band photometric observations}
\label{sec:v_band_observations}

\begin{table*}
\centering          
\begin{tabular}{l c c c c c}
Survey & \# of observations & Median magnitude [mag]& Date range & Mean cadence [days] & Median error [mag] \\
\hline
ASAS & 387 & 8.66 & 2000-12-08 -- 2009-03-28 & 7.85 & 0.025 \\
OMC & 1654 & 8.62 & 2003-01-29 -- 2019-01-27 & 3.53 & 0.013 \\
ASAS-SN & 632 & 8.69 & 2016-02-05 -- 2018-07-22 & 1.42 & 0.010 \\

  \hline
\end{tabular}
\caption{The number of observations, median magnitude, date range, mean cadence, and median error for the V-band photometric data of GG Car for each survey used in this study.}
\label{tab:v_band_observations}      
\end{table*}

V-band photometric data of GG Car is available from the All Sky Automated Survey (ASAS, \citealt{Pojmanski2002TheHemisphere, Pojmanski2004TheSurvey}), the Optical Monitoring Camera aboard the INTEGRAL satellite (OMC, \citealt{Mas-Hesse2003OMC:INTEGRAL}), and the All Sky Automated Survey for Supernovae (ASAS-SN, \citealt{Shappee2014THE2617, Kochanek2017TheV1.0}). Each of these surveys use standard Johnson V-filters, centred at 550\,nm and with a full width half maximum of 88\,nm. Further details of the V-band observations used in this study for each survey are given in Table \ref{tab:v_band_observations}. The median magnitudes between the surveys, listed in this table, are consistent with each other, given that GG Car is a variable between magnitudes $\sim$8.4--9.0 and each survey has a different observation window and phase coverage.\\

The ASAS-SN lightcurve for GG Car is calculated using the project's \texttt{Sky Patrol} feature, where the user can enter any celestial coordinate and a light curve is returned. Ostensibly the saturation limit for the ASAS-SN cameras is around 10th magnitude; however, corrections are done to the photometry automatically for brighter sources, such as GG Car. Occasionally the correction of the data fails, and therefore spurious data need to be removed manually. ASAS-SN data are also affected by the phase of the moon; we model this as a sinusoidal effect in brightness with a period of 29.51 days, and this model is subtracted from the data.\footnote{For further examples in failures of ASAS-SN flux correction due to saturation, see figure 9 in \cite{Kochanek2017TheV1.0} and example 5 in \url{http://www.astronomy.ohio-state.edu/~assassin/public/examples.shtml}. Example 2 of the same web page shows the effect of the phase of the Moon on the photometry.} \\

\subsection{Global Jet Watch spectroscopy}

The Global Jet Watch (GJW) has been collecting mid-resolution (R\,$\sim$\,4\,000) optical spectroscopic data on a variety of objects, including GG Car which it has been observing since early 2015. GJW is an array of five telescopes, separated in longitude, which take optical spectra from $\sim$\,5\,800\,--\,8\,400\,\AA. The spectra are reduced with a bespoke data reduction pipeline making use of dark and flat field calibration exposures, and the wavelengths are calculated using Thorium Krypton calibration frames. The spectra are barycentric corrected using heliocentric velocities calculated with the \texttt{barycorrpy} package \citep{Kanodia2018Pythonbarycorrpy}. In this study, all spectra are normalised by the local continuum.\\

\begin{table} 
\centering          
\begin{tabular}{ l l}
\hline \hline
\\
    Exposure time [seconds] & \# of spectra \\
    \hline \\
    100 & 596 \\
    1000 & 305 \\
    3000 & 416 \\

\end{tabular}
\caption{The number of GJW spectra used in this study.}
\label{tab:exposures}      
\end{table}

 GJW's wavelength range is particularly well suited to study a number of GG Car's emission features, such as the dominant H-alpha line at 6563\,\AA, He\,I lines at 5875, 6678, and 7065\,\AA, and an abundance of permitted Si\,II and Fe\,II emission lines. Our observations of GG Car have exposure times of either 100, 1000, or 3000 seconds. This range of exposure times is optimised for observing the bright H-alpha line, with the 100 second exposures, and the metal and He\,I lines, which require at least 1000\,second exposures to achieve acceptable signal-to-noise. Table \ref{tab:exposures} lists the number of observations at each exposure time, and Figure \ref{fig:n_spectra_by_phase} in the appendix displays the distribution of these observations over the orbital period of the binary.\\

\subsection{FEROS spectroscopy}
\label{sec:feros}
The Fiber-fed Extended Range Optical Spectrograph (FEROS, \citealt{Kaufer1999CommissioningLa-Silla.}) is a high-resolution echelle spectrograph, located at the European Southern Observatory (ESO) at La Silla, Chile. Before October 2002, the spectrograph was used with the ESO 1.52m telescope; since then it has been used with the 2.2m MPG/ESO telescope. FEROS has a wavelength range of 3600\,-\,9200\,\AA\ and $R$\,$\sim$\,$48\,000$. The FEROS spectra used in this study have been reduced by ESO using the standard FEROS reduction pipeline. \\

The publicly available FEROS data of GG Car span three epochs:  December 1998, May 2015 and November 2015. Table \ref{tab:feros_observations} lists the exact dates and times of the observations. The FEROS data do not have satisfactory phase coverage to undertake time-series analysis, as is shown in Figure \ref{fig:feros_phase_coverage} in the appendices. However, the high-spectral resolution observations may be used to complement the high-temporal resolution GJW data by revealing the fine structure of the emission lines' profiles, and thereby revealing the formation regions of the lines, allowing us to choose unblended lines to investigate, and allowing us to make estimates of the terminal wind velocity. \\

\section{Revised luminosity and mass of the B[\,e\,] supergiant primary of GG Car}
\label{sec:gaia_luminosity}
\textit{Gaia}'s Data Release 2 \citep{Prusti2016TheMission, Brown2018Gaia2} published a parallax to GG Car of $0.2997 \pm 0.0303$\, mas; using the Uniform Distance Prior method of \cite{Luri2018Gaia2} to convert parallax to a distance, this indicates a distance to GG Car of $3.4^{+0.7}_{-0.5}$\,kpc and a distance modulus of $12.66^{+ 0.41}_{-0.34}$\,mag. \\

\textcolor{black}{We show in Section \ref{sec:enhanced_mass_transfer} that the photometric variations at the orbital period are due to luminosity increases at periastron due to mass transfer between the components, and in Section \ref{sec:secondary_mass} we show that the luminosity of the secondary is expected to be negligible compared to the primary.} Therefore, we assume that the V-band brightness of the system when it is at photometric minimum, at apastron, is solely dominated by the intrinsic brightness of the \sgBeshorthand\ primary. We therefore assign the primary an intrinsic V-band brightness of $m_V = 8.8 \pm 0.1$\,mag (Figure \ref{fig:v_band_31_fold}). We can find the absolute V-band magnitude of the primary, $M_V$, by
\begin{equation}
    M_V = m_V - 5 \log\left( \frac{d}{10\,{\rm pc}}\right) - A_V = m_V - \mu - A_V,
\end{equation}

\noindent where $d$ is the distance to GG Car in parsecs, $A_V$ is the interstellar extinction in the V-band, and $\mu$ is the distance modulus. \\

$A_V$ and the $B-V$ excess, $E_{B-V}$, are proportional to each other, such that

\begin{equation}
    R_V = \frac{A_V}{E_{B-V}},
\end{equation}

\noindent where $R_V$ is the ratio of absolute to relative extinction. It is often assumed that within the Galaxy $R_V \sim 3.2$; however, $R_V$ is highly dependent on the properties of the interstellar medium (ISM) along the line of sight to the target (e.g. \citealt{Majaess2016ConstrainingSurvey}). GG Car's Galactic coordinates (289.133, -00.651) places it within the Galactic plane, where the ISM densities will be greatest. Furthermore, the star-forming Carina nebula has often been found to have an anomalously high $R_V$ (e.g. \citealt{Feinstein1973A16., Herbst1976OnNebula., Smith1987An3372, Tapia1988ThePhotometry, Patriarchi2003ACatalogue, Hur2012DISTANCE16, Majaess2016ConstrainingSurvey}) and extinction which can be significantly more than 5\,mag in the V-band (e.g. \citealt{Preibisch2014TheSurvey, Damiani2017GaiaNebula}). In a study of the dependence of $R_V$ on the Galactic longitude, $l$, for O-stars, \cite{Majaess2016ConstrainingSurvey} found a sharp peak in $R_V$ at $l \sim 290^\circ$, almost exactly at the Galactic longitude of GG Car. The cited studies find $R_V$ in the range 4.4\,--\,4.8 toward the Carina region, therefore we adopt $R_V=4.6\pm0.2$ for the ISM along the line of sight to GG Car for our calculation of its primary's luminosity. \cite{Brandi1987} and \cite{Marchiano2012} found $E_{B-V}$ to be $0.52\pm0.04$\,mag and $0.51\pm0.15$\,mag, respectively; therefore, we adopt a total visible extinction $A_V = 2.35\pm0.21$\,mag and find $M_V = -6.21\pm0.44$\,mag.\\

We calculate the bolometric correction, $BC_V$, of GG Car to be $-2.21\pm0.20$\,mag, using the \cite{Flower1996TransformationsCorrections} regime for a star of $\log (T_{\rm eff}) > 3.90$, which was corrected in \cite{Torres2010ONSTARS}, assuming an effective temperature $T_{\rm eff}= 23000\pm2000$\,K \citep{Marchiano2012}. Therefore, the bolometric magnitude of the primary $M_{\rm bol} = M_V + BC_V = -8.42\pm0.49$\,mag. The luminosity of the primary can then be calculated using the relation
\begin{equation}
    M_{\rm bol} - M_{\rm bol, \odot} = -2.5 \log \left(\frac{L}{L_\odot}\right),
\end{equation}

\noindent where $M_{\rm bol, \odot}$ is the solar bolometric magnitude, and $L_\odot$ is the solar luminosity. To be consistent with the $BC_V$ calculated using \cite{Flower1996TransformationsCorrections} and \cite{Torres2010ONSTARS}, we adopt $M_{\rm bol, \odot} = 4.73$\,mag. Therefore the luminosity of the primary of GG Car is $\log(L_{\rm pr}/L_\odot) = 5.26\pm0.19$. This corresponds to $L_{\rm pr} = 1.8^{+1.0}_{-0.7} \times 10^5 \, L_\odot$. The effective temperature and updated luminosity imply a stellar radius $R_{\rm pr} = 27^{+9}_{-7}\,R_\odot$. \textcolor{black}{Our determination of $L_{\rm pr}$ agrees well with \cite{Krtickova2018AnEnvelopes}, who found that the \sgBeshorthand s of the LMC and the SMC, with their well-defined distances, had remarkably uniform luminosities of $\sim (1.9 \pm 0.4) \times 10^5\,L_\odot$.}\\

 \begin{figure} 
  \centering
    \includegraphics[width=0.5\textwidth]{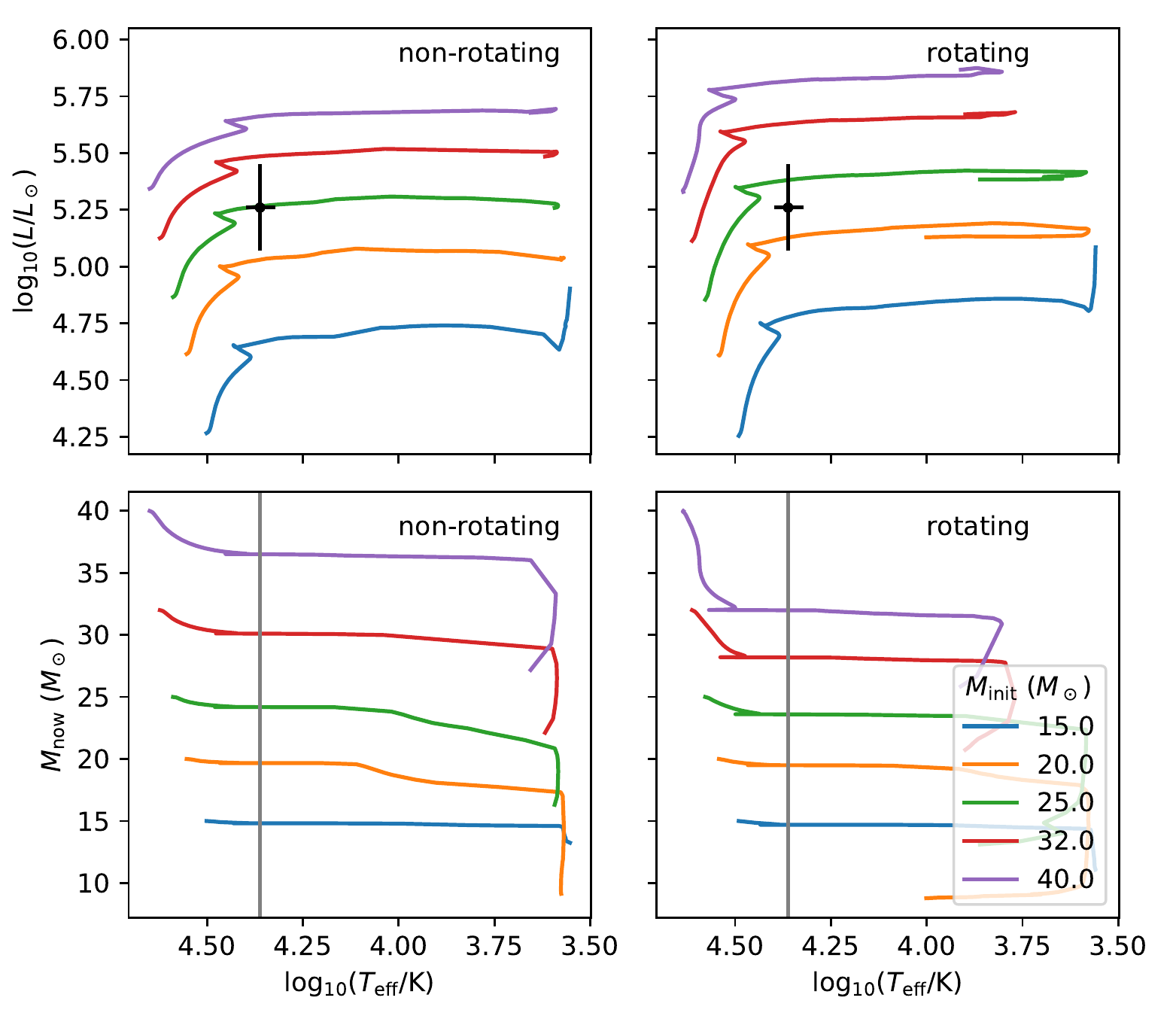}
    \caption{Stellar evolutionary models of luminosity, $L$, and stellar mass, $M_{\rm now}$, against effective temperature, $T_{\rm eff}$, of \protect\cite{Ekstrom2012GridsRotation} for varying initial stellar mass, $M_{\rm init}$. The left-hand panels shows the tracks for the non-rotating cases, and the right-hand panels shows the rotating cases. The position of GG Car on the H-R diagram is denoted by a black point. The $T_{\rm eff}$ of the primary of GG Car is denoted by the grey, vertical line in the $M_{\rm now}$ panels. Only data where the ratio of the abundances of the Carbon isotopes $^{12}{\rm C} / ^{13}{\rm C} > 5$ are plotted. The legend in the bottom-right panel gives $M_{\rm init}$ for the tracks in all panels.}
    \label{fig:evolutionary_tracks}
\end{figure}

\textcolor{black}{The direct observation of the primary star's mass is precluded by the thick circumstellar envelope of the system. Therefore, to estimate the mass of GG Car's primary, we must compare its position on the Hertzsprung-Russell (H-R) diagram to stellar evolution models.} We choose to compare its measured luminosity and effective temperature to the grid of stellar evolution models of \cite{Ekstrom2012GridsRotation}, who calculated stellar evolution tracks for stars with initial masses from 0.8 to 120\,$M_\odot$ at a solar metallicity of $Z$\,=\,0.014; this is similar to the observed metallicity towards the Carina region \citep{Spina2017TheClusters}. The models were calculated for both initially rotating and non-rotating stellar models, with the rotating stellar models having an initial rotational velocity $\sim 0.4\, v_{\rm crit}$, where $v_{\rm crit}$ is the critical break up velocity of the star. Figure \ref{fig:evolutionary_tracks}, top two panels, displays the position of GG Car's primary in the H-R diagram, along with the evolutionary tracks at solar metallicity for a suitable range of initial masses of \cite{Ekstrom2012GridsRotation}, in the rotating (right panel) and non-rotating (left panel) cases. From its position on the H-R diagram relative to the evolutionary tracks, we can estimate the initial mass of the primary. We estimate $20\,M_\odot < M_{\rm init} < 32\,M_\odot$, where $M_{\rm init}$ is the initial mass of the primary of GG Car. The bottom two panels of Figure \ref{fig:evolutionary_tracks} display the evolution of the mass of the stars in the \cite{Ekstrom2012GridsRotation} models, and the current $T_{\rm eff}$ of the primary of GG Car; from the initial mass estimate from the top two panels and the evolution of the mass together, we can deduce that $20\,M_\odot < M_{\rm now} < 28\,M_\odot$, where $M_{\rm now}$ is the current mass of the primary star in GG Car. We therefore deduce the primary star's mass $M_{\rm pr} = 24\pm4\,M_\odot$, and list the updated stellar parameters in Table \ref{tab:ggcar_parameters}. \textcolor{black}{It must be noted that this deduced mass is model dependent using the single-star evolution models of \cite{Ekstrom2012GridsRotation}, and that if there has been significant binary interaction in GG Car's history that the primary has regained thermal equilibrium (see Section \ref{sec:binary_evolution}).}\\

\begin{table}
\centering          
\begin{tabular}{ l l}
\hline \hline
\\
    $d$ & $3.4^{+0.7}_{-0.5}$\,kpc\\
   $M_{\rm pr}$  & $24\pm4\,M_\odot$ \\ 
   $T_{\rm eff}$ & $23\,000 \pm 2000$\,K \\
   $L_{\rm pr}$ & $1.8^{+1.0}_{-0.7}\times10^5\,L_\odot$ \\
   $R_{\rm pr}$ & $27^{+9}_{-7}\,R_\odot$ \\
\end{tabular}
\caption{\textit{Gaia} distance, $d$, and stellar parameters of the primary in GG Car, where $M_{\rm pr}$ is the mass of the primary, $T_{\rm eff}$ is the effective temperature of the primary, $L_{\rm pr}$ is the luminosity of the primary, and $R_{\rm pr}$ is the radius of the primary. $T_{\rm eff}$ is taken from \protect\cite{Marchiano2012}.} 
\label{tab:ggcar_parameters}      
\end{table}

\section{The variability of GG Car along the orbital period}
\label{sec:variability}

\subsection{Photometric variability}
\label{sec:photometric_variability}

\begin{figure} 
  \centering
    \includegraphics[width=0.5\textwidth]{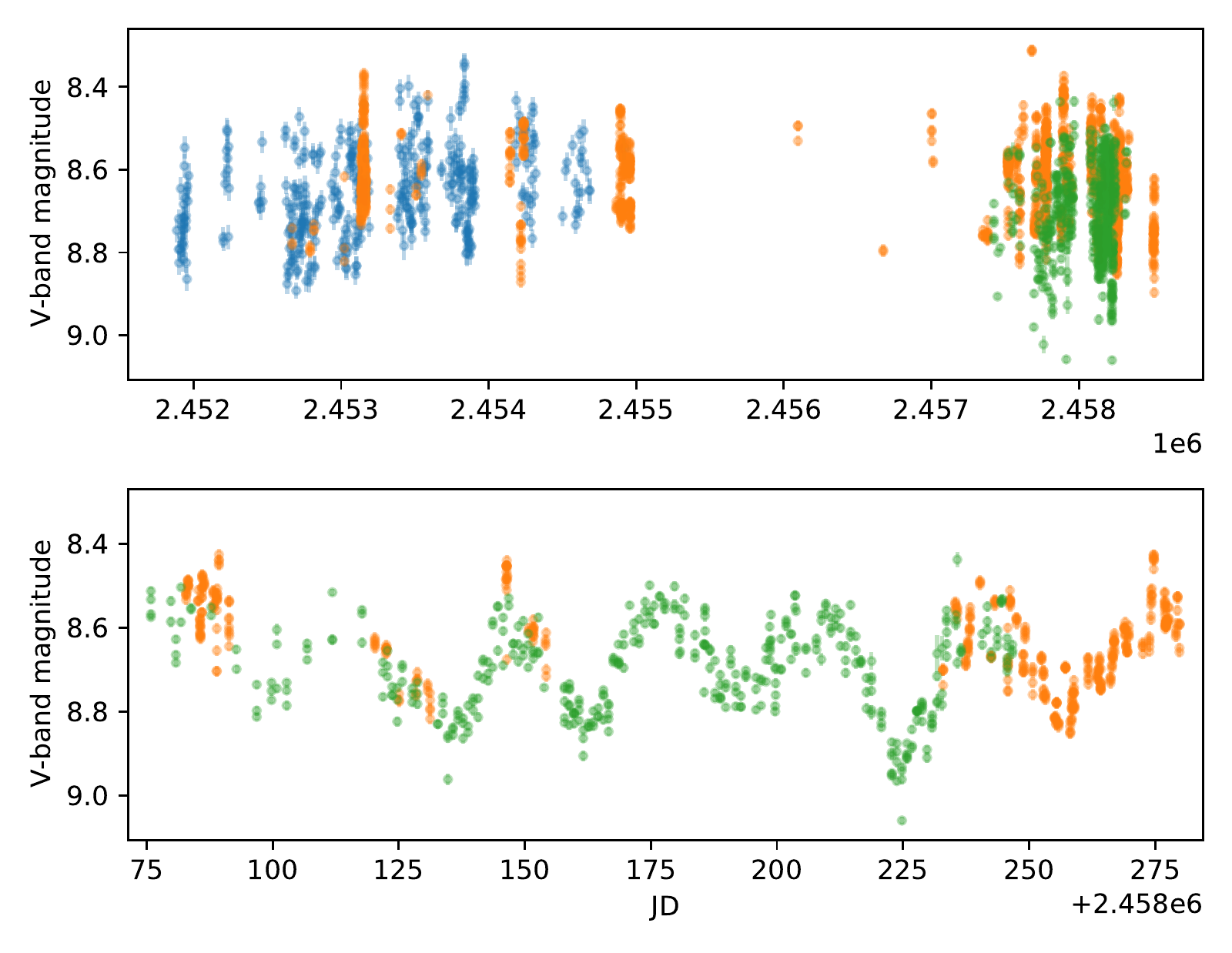}
    \caption{V-band photometry of GG Car from ASAS (blue), OMC (orange), and ASAS-SN (green). The top panel displays all available data, and the bottom panel displays a few finely-sampled orbital periods within the ASAS-SN and OMC data sets.}
    \label{fig:gg_car_v_band_photometry}
\end{figure}

Figure \ref{fig:gg_car_v_band_photometry} displays the V-band photometry of GG Car from ASAS, ASAS-SN, and OMC. The variability is dominated by the $\sim$31-day period of the binary, as can be seen in the bottom panel. There is also some longer term variability present in the data, and noticeable scatter along the lightcurve. As these are mostly ground-based experiments within much larger monitoring programmes, these data have, at best, only a few observations per day.\\

\begin{figure} 
  \centering
    \includegraphics[width=0.5\textwidth]{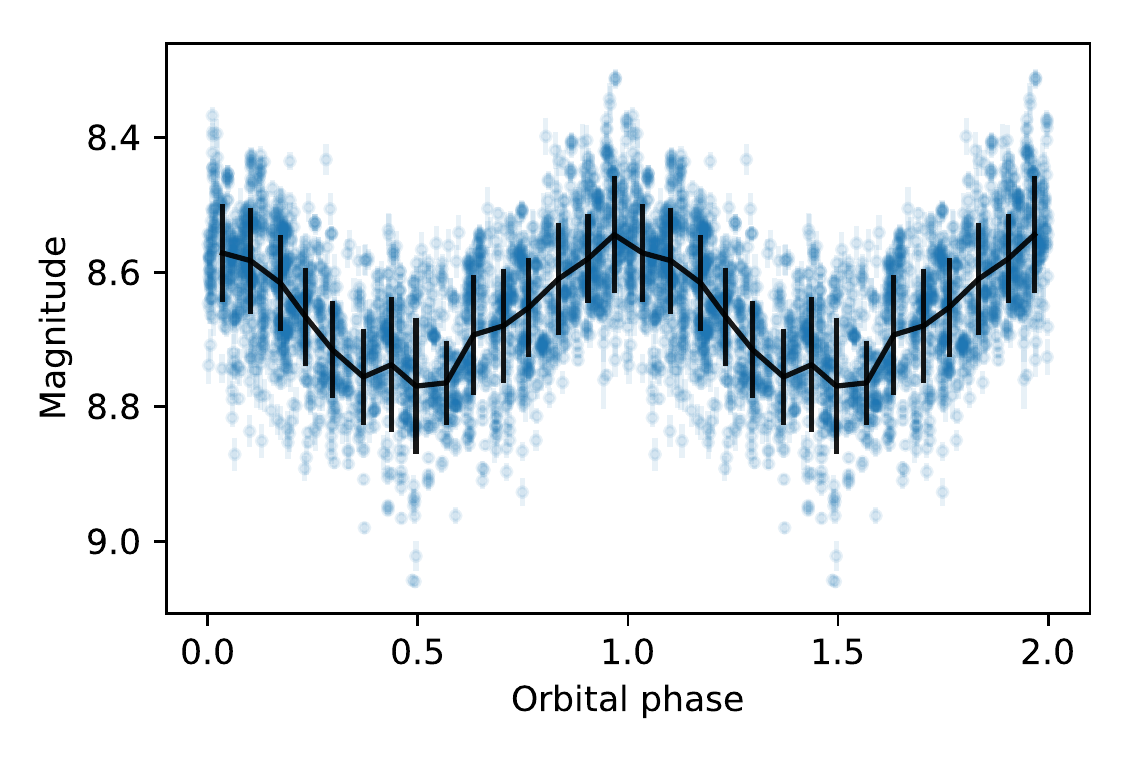}
    \caption{V-band photometry of GG Car, folded by the orbital phase of the binary using the ephemeris of Equation \ref{eq:ephemeris}. The black line indicates the mean magnitude in 15 phase bins, and the error bar shows the standard deviation in the bin. The data are repeated for two orbital cycles.}
    \label{fig:v_band_31_fold}
\end{figure}

Figure \ref{fig:v_band_31_fold}, displays the V-band photometry of GG Car folded by the orbital period using the orbital ephemeris of Equation \ref{eq:ephemeris}, which is determined in Section \ref{sec:determining_orbital_solution}. The black line shows the average magnitude in 15 phase bins, and the errorbars show the standard deviation of the magnitudes in the bin. There is a lot of scatter in the system's light curve, as is shown by the errorbars, with the average standard deviations in the bins of 0.078\,mag. The mean brightness is continuous, with only one photometric minimum per orbital period and no sign of eclipsing. At photometric minimum the standard deviation of the brightness is slightly higher than at other phases.\\

\begin{figure} 
  \centering
    \includegraphics[width=0.5\textwidth]{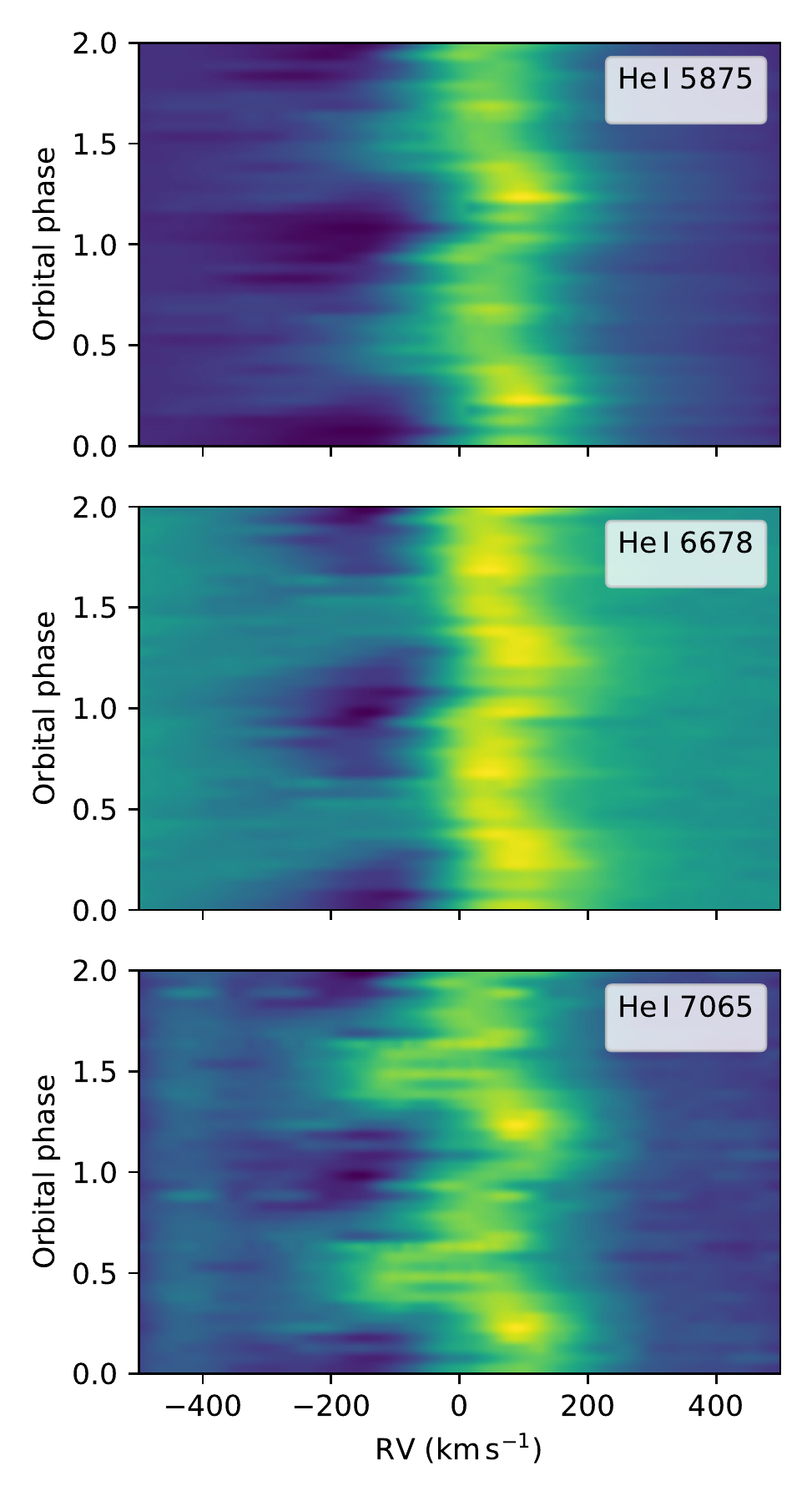}
    \caption{Trailed spectra of GG Car's visible He\,I lines, as observed by GJW. The spectra are split into 20 bins by orbital phase, and the average spectrum of the continuum-normalised spectra in each bin is displayed. Yellow is higher intensity, blue is lower intensity.}
    \label{fig:he_trailed_spectra}
\end{figure}

A cause of the large amount of scatter is that, as can be seen in Figure \ref{fig:gg_car_v_band_photometry}, the photometric variations for each 31-day orbital cycle are clearly not identical, but the profiles are stochastic and vary in shape and depth. In addition, a shorter-period and very long-term non-periodic variations are present in the system which are not subtracted from the data; the shorter-period variations are discussed in detail in an upcoming study. Fitting the long-period variations with polynomials and subtracting them from the data only minimally affects the scatter observed, \textcolor{black}{and therefore is not implemented in our analysis}.\\

The black, sliding average, magnitude in Figure \ref{fig:v_band_31_fold} ranges from 8.55--8.77 magnitudes, indicating an average V-band magnitude change of 0.22 over the orbital period; this corresponds to a flux change of 21.9\% over the binary's orbit. As mentioned, the V-band variability over the orbital period is stochastic, so this value is the average brightness change in an orbital period, but this value may vary for successive orbits. The presence of one photometric maximum and one minimum per orbital period matches previous photometric studies. However, we do not find evidence of the ``glitch'' observed by \cite{Gosset1985} at photometric maximum. \\

\begin{figure} 
  \centering
    \includegraphics[width=0.5\textwidth]{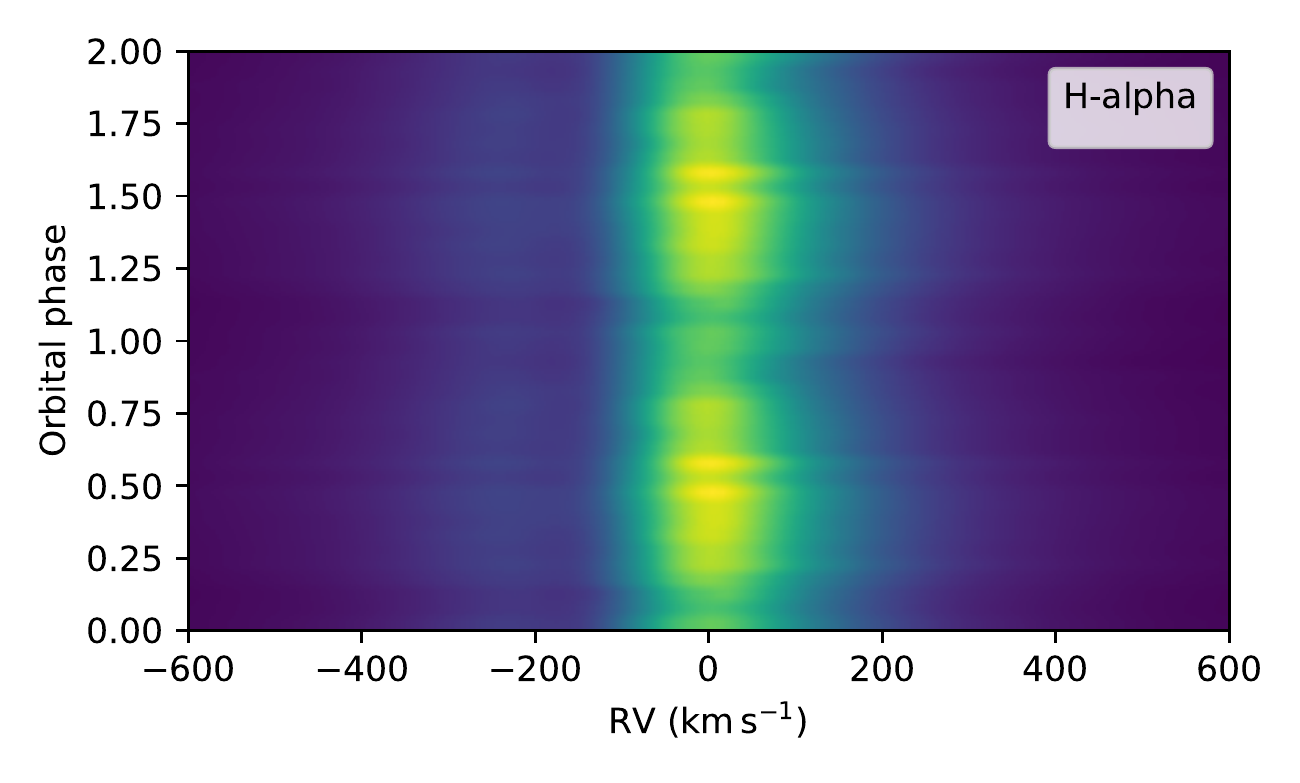}
    \caption{Same as Figure \ref{fig:he_trailed_spectra}, except for GG Car's H-alpha complex.}
    \label{fig:halpha_trailed_spectra}
\end{figure}

\subsection{Spectroscopic variability}
\label{sec:spectroscopic_variability}
In the GJW spectra, the most notably variable emission lines along the orbital period are He\,I 5875, 6678, and 7065, which usually exhibit P-Cygni profiles, indicating mass loss. However, these He\,I profiles are highly dynamic and may vary between pure emission and pure absorption. In these mid-resolution spectra, there is usually only one (or no) discernible absorption feature in the He\,I lines, which we interpret as a blend of the multiple wind features expected for \sgBeshorthand s \citep{Zickgraf1985, Oudmaijer1998TheStars}. Figure \ref{fig:he_trailed_spectra} displays trailed spectra of the He\,I lines, ordered by the orbital phase of the binary. The blueshifted absorption is deepest around phase 0 and shallowest and often non-existent around phase 0.5. For He\,I 5875 and 6678, the emission follows a simple continuous variation where it is most redshifted around phase 0.15, and most blueshifted at phase 0.78. He\,I 7065, on the other hand, hosts similar variations to 5875 and 6678, but gains a blueshifted emission component between phase 0.3 to 0.6, at around $-$120\,\kms. This extra component is unclear in individual spectra, but becomes apparent when successive spectra are taken in aggregate in the trailed spectra.\\

Here we quantify the radial velocity (RV) variability of the centres of the He\,I emission. It has been shown that Gaussian fitting is a robust and simple method of identifying the centres of emission lines that originate in winds (e.g. \citealt{Blundell2007FluctuationsTimescales, Grant2020Uncovering140}). Therefore, we fit the He\,I 5875 and 6678 lines in each GJW spectrum using a model consisting of one positive Gaussian and one negative Gaussian component, via a least-squares optimisation method implemented in \texttt{scipy} \citep{Virtanen2020SciPyPython}. The centroid of the positive Gaussian component is taken as the RV of the emission. He\,I 7065 is fitted in a similar manner, except two positive and one negative Gaussians are used; the centroid of the more redshifted positive Gaussian is then taken as the RV of the emission.\\

In comparison to the He\,I lines, the H-alpha complex in GG Car is remarkably constant over the orbital period, as is shown in Figure \ref{fig:halpha_trailed_spectra}. There are no large RV changes and the blueshifted absorption persists at the same RV of around $-$160\,\kms\ for all phases without any measurable RV variability. In Figure \ref{fig:halpha_trailed_spectra}, it appears as the intensity of the peak changes as a function of the orbital phase; however, this is due to the brightness variations along the orbital period. Since the spectra are normalised by the local continuum, around phase 0, when the system is brightest, the continuum is stronger which makes the H-alpha line appear dimmer in comparison. Conversely, at phase 0.5 the continuum is less bright, meaning the H-alpha complex appears brighter in comparison. This effect makes the equivalent widths of all emission lines a product of the continuum level rather than the emission line strengths. The wings of the H-alpha line, unlike the absorption or peak, do noticeably vary in radial velocity over the orbital period. Therefore, to extract the RV variations of the wings, we fit the H-alpha complex with multiple Gaussian components, including a negative Gaussian to account for the blueshifted absorption. The widest Gaussian component is then taken to be the wings of the H-alpha line.\\

\begin{figure} 
  \centering
    \includegraphics[width=0.5\textwidth]{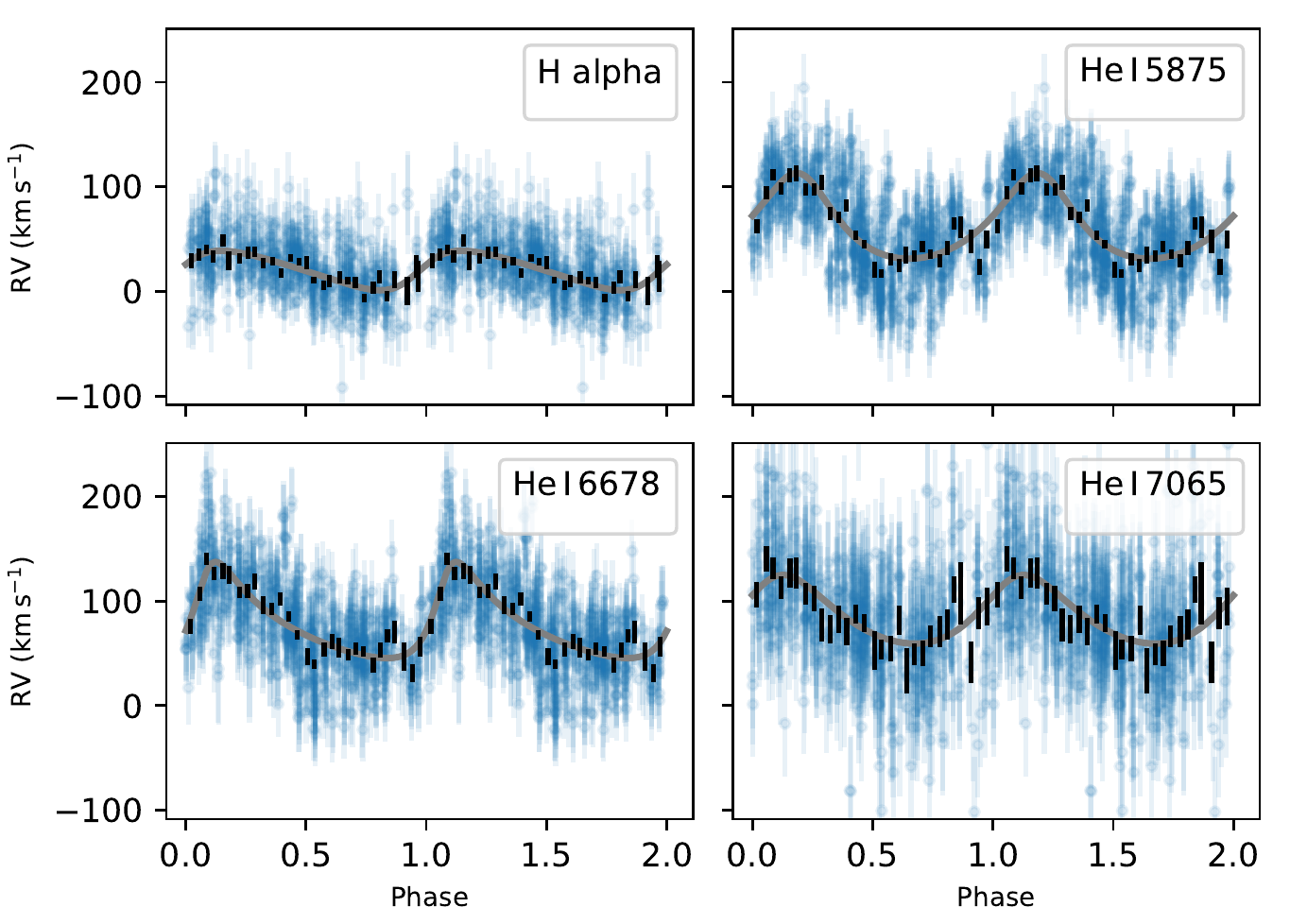}
    \caption{Radial velocity variations of the emission components of the visible He\,I lines at the 31-day orbital period, as observed by the GJW. Orbital solution fits are plotted as solid grey lines. The orbital models from the short-period variations are subtracted from the data, for clarity. The data are split into 30 phase bins for each line species and black error bars indicate the weighted mean and standard error of the weighted mean for the radial velocities in each bin. The error bars are both the errors from the Gaussian fitting routine and the fitted jitter, added in quadrature.}
    \label{fig:he_31_rv_variability}
\end{figure}

Orbital RV solutions are fitted to the RV data for each line separately at both the orbital- and 1.583-day periods simultaneously, since the amplitude of the short-period variations are non-negligible for the He\,I lines. We fit for the amplitude $K$, the eccentricity $e$, the argument of periapsis $\omega$, and the phase of periastron $M_0$ for each period. We also fit for jitter, $j$, which models the width of Gaussian-distributed scatter of the data over the orbital period. We also fit the systemic velocity $v_0$ separately for each line. We fit the RV variations for each emission line separately by maximising the log-likelihood function, using the Monte Carlo Markov Chain (MCMC) algorithm \texttt{emcee} \citep{Foreman-Mackey2012Emcee:Hammer}. The log-likelihood for a set of parameters, $\theta$, given $N$ RV data points, $D$, with uncertainty $\sigma$ is given by
\begin{multline}
\label{eq:log_like}
\begin{split}
    \text{ln} \, P (\theta \mid D, \, \sigma) = - \frac{1}{2} \sum_{i=0}^N \left[\frac{\left(D_i - v_{\rm kep}(\theta_{\rm orb}) - v_{\rm kep}(\theta_{1.583}) - v_0\right)^2}{\sigma_i^2 + j^2}  \right.\\ \left.+ \text{ln}\left(2\pi (\sigma_i^2 + j^2)\right)\right],
\end{split}
\end{multline}
\noindent where $\theta_{\rm orb}$ and $\theta_{1.583}$ are the orbital parameters for the long- and the short- period respectively, and $v_{\rm kep}$ is the Keplerian velocity calculated for a set of orbital parameters. The fitted parameters, $\theta$, encodes $\theta_{\rm orb}$, $\theta_{1.583}$, $v_0$, and $j$. The uncertainties, $\sigma$, are taken from the least-squares fitting algorithm of the Gaussian fitting routines. We run the MCMC for 2000 burn iterations, then 2000 sampling iterations with 128 walkers. The fitted parameters are taken to be the median value of the Monte Carlo samples, and the uncertainty of the fitted parameters is taken as the difference of the median with the 16th and 84th percentiles. Table \ref{tab:orbital_solutions} in the appendix list the fitted orbital parameters of the 31-day variations for to the He\,I and H-alpha emission lines. For more details of the 1.583-day variations we refer the reader to our upcoming study, where the short-period will be discussed. \\

Figure \ref{fig:he_31_rv_variability} displays the RV variations of the H-alpha wings and the He\,I emission components at the orbital period, and the resulting fits for each period are plotted over the data. The data are binned by phase into 30 bins, and the weighted mean and standard error on the weighted mean in each bin are given by the black error bars. The RV variations have different amplitudes and profiles for each line species, with H-alpha having the subtlest variations. The RV variations of He\,I 6678 and 7065 have very similar profiles, with He\,I 5875 deviating slightly around phase 0.5, becoming more blueshifted. This is due to a second component arising in this line at that phase, similar to He\,I 7065; it can be seen very faintly in the trailed spectra in Figure \ref{fig:he_trailed_spectra}, top panel. However, the second component is not as overwhelming in the 5875 line as the 7065 line, therefore attempting to fit this intermittent component in He\,I 5875 generally adds error to the fitting routine as the line becomes over-fit in these mid-resolution spectra.\\

\begin{figure} 
  \centering
    \includegraphics[width=0.5\textwidth]{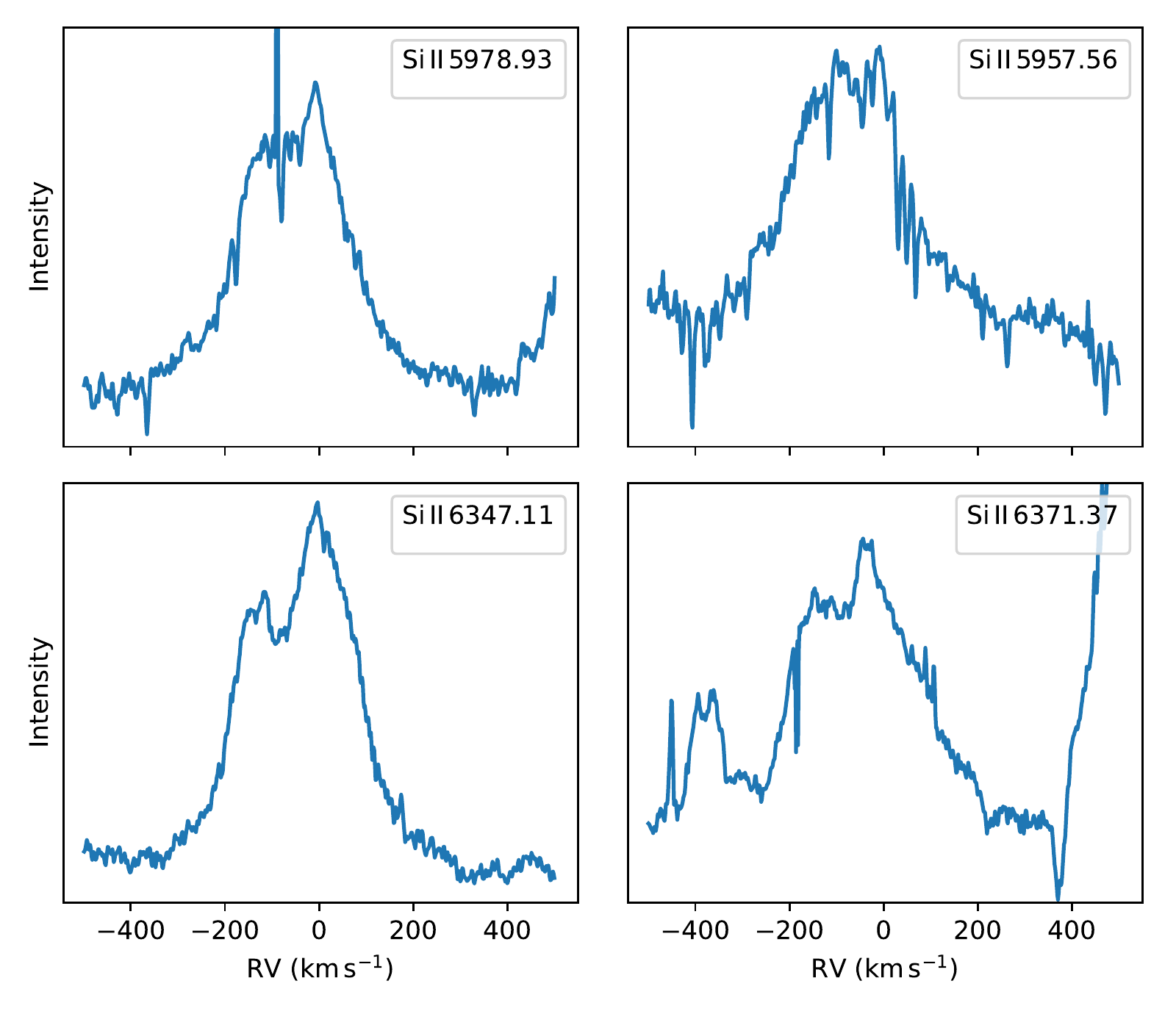}
    \caption{Average profiles of the Si\,II lines used in this study, as observed by FEROS.}
    \label{fig:feros_si_lines}
\end{figure}

\begin{figure} 
  \centering
    \includegraphics[width=0.5\textwidth]{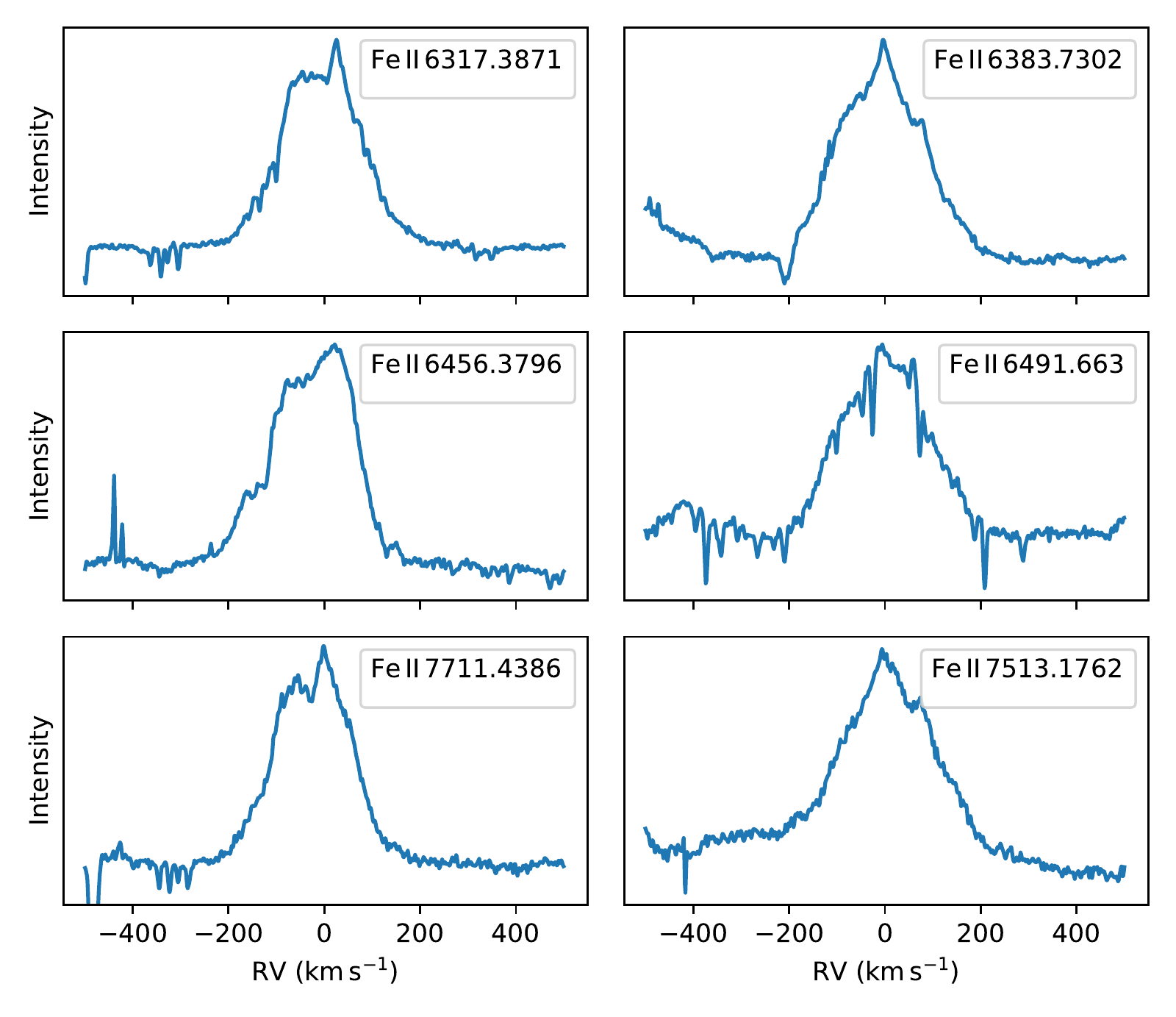}
    \caption{Average profiles of the Fe\,II lines used in this study, as observed by FEROS.}
    \label{fig:feros_fe_lines}
\end{figure}

Several Si\,II lines and Fe\,II lines also display RV variability at the orbital period. Due to the wealth of Fe\,II lines in the visible wavelength range, we take care to investigate lines which are unblended. The choosing of lines is done with the aid of the FEROS data, as the high resolution data allow us to clearly discriminate between emission lines. Figure \ref{fig:feros_si_lines} displays the average profiles of the Si\,II lines chosen for this study. Likewise, Figure \ref{fig:feros_fe_lines} displays the average profiles of the chosen Fe\,II lines which display 31-day variability. Certain Fe\,II lines showed no RV variability in their centroids; two such lines were Fe\,II 5991.4 and 6432.7, which are shown in Figure  \ref{fig:feros_fe_disk_lines} in the appendix. In the high-resolution FEROS spectra, these lines exhibit distinctive double peaked disk profiles with a peak-to-peak splitting of $\sim$160\,km\,s$^{-1}$; therefore, due to their profiles and lack of RV variability, we ascribe them to the circumbinary disk of the system, and study them in detail an upcoming publication. \\

\begin{figure} 
  \centering
    \includegraphics[width=0.5\textwidth]{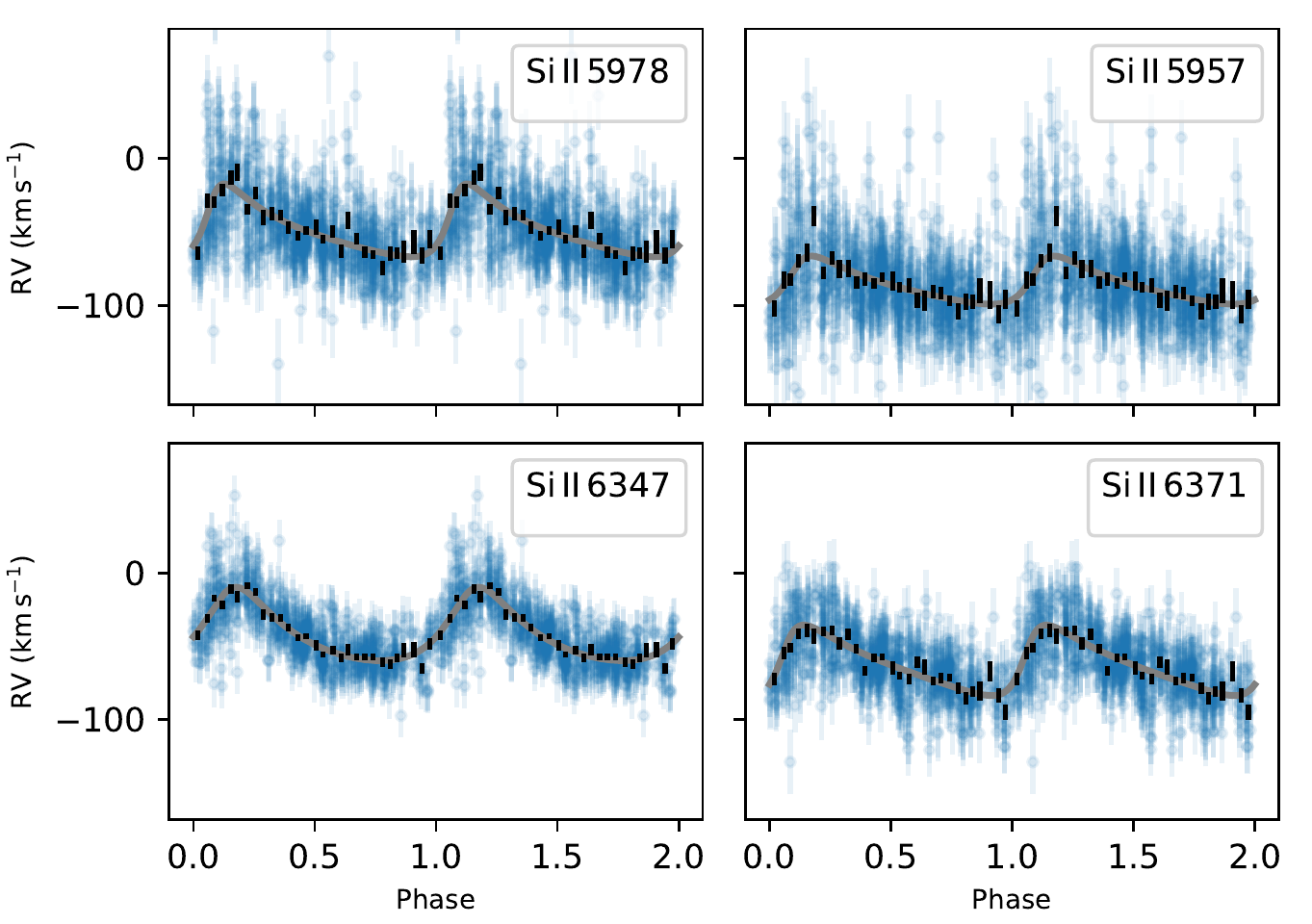}
    \caption{Same as Figure \ref{fig:he_31_rv_variability}, except for the four Si\,II lines.}
    \label{fig:si_31_rv_variability}
\end{figure}

\begin{figure} 
  \centering
    \includegraphics[width=0.5\textwidth]{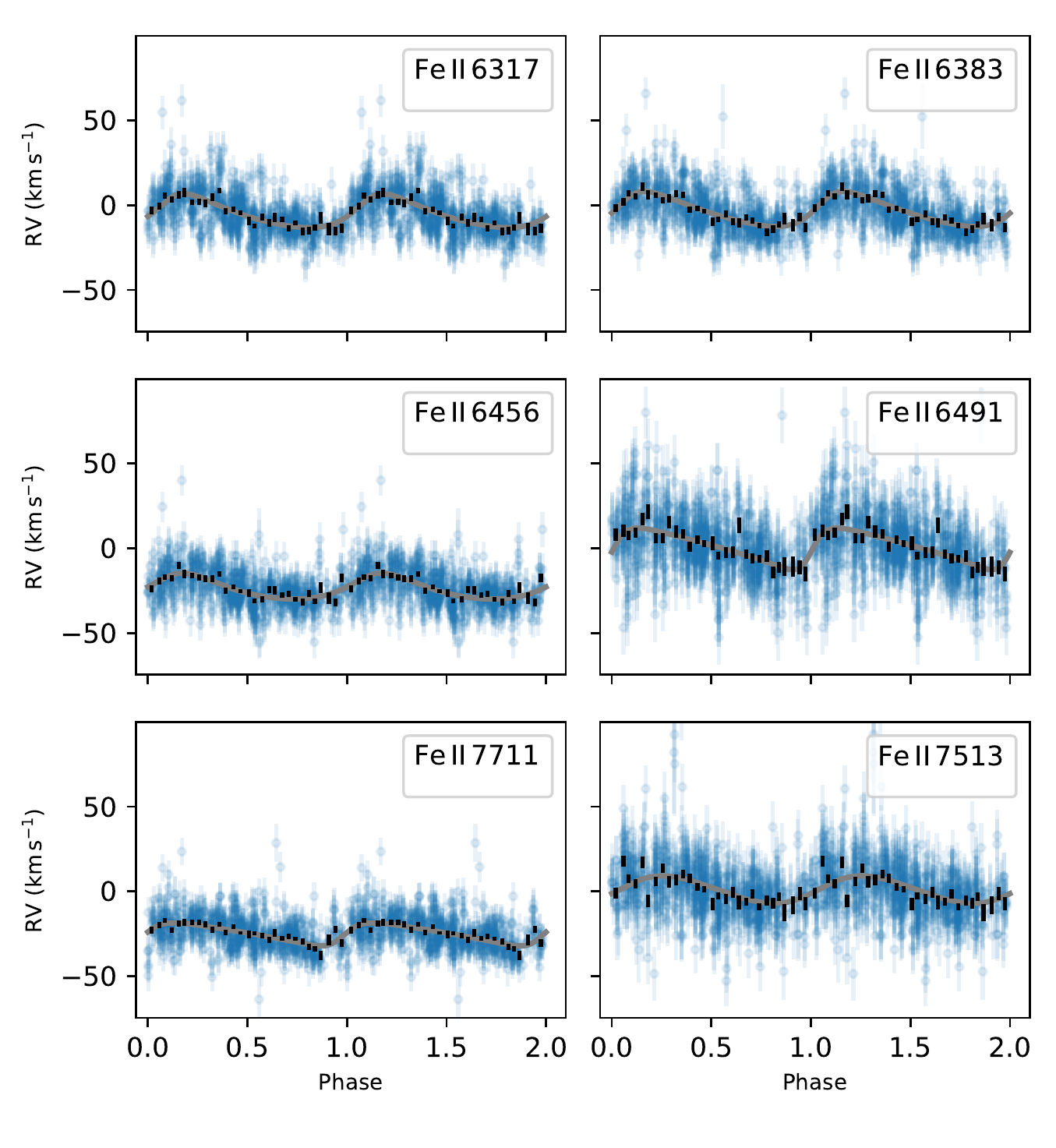}
    \caption{Same as Figure \ref{fig:he_31_rv_variability}, except for the Fe\,II lines.}
    \label{fig:fe_31_rv_variability}
\end{figure}

The RVs of the Si\,II and Fe\,II lines were calculated by fitting single positive Gaussian profiles to them in the GJW data, since they do not suffer significant absorption and the structure of each line is unresolved. The centroid of the Gaussian is then taken as the RV. Figures \ref{fig:si_31_rv_variability} and \ref{fig:fe_31_rv_variability} display the RV variations for the Si\,II and Fe\,II lines, respectively. Orbital solution models were fitted using the same method as for the He\,I and H-alpha emission lines, and the resulting fits are also presented in Table \ref{tab:orbital_solutions}. The Si\,II lines tend to be more blueshifted than the systemic velocity of the system, and have smaller RV variations than the He\,I lines. The systemic velocities of the Si\,II lines vary for the different species. The Fe\,II lines, on the other hand, have much more consistent systemic velocities. In these lines, there appear to be two families of systemic velocities, those around $\sim$0\,\kms, and those around $-$25\,\kms. This matches \cite{Gosset1985}, who also found that the systemic velocities interpreted from Fe\,II lines depended on the multiplet observed. \\\\

\begin{figure} 
  \centering
    \includegraphics[width=0.5\textwidth]{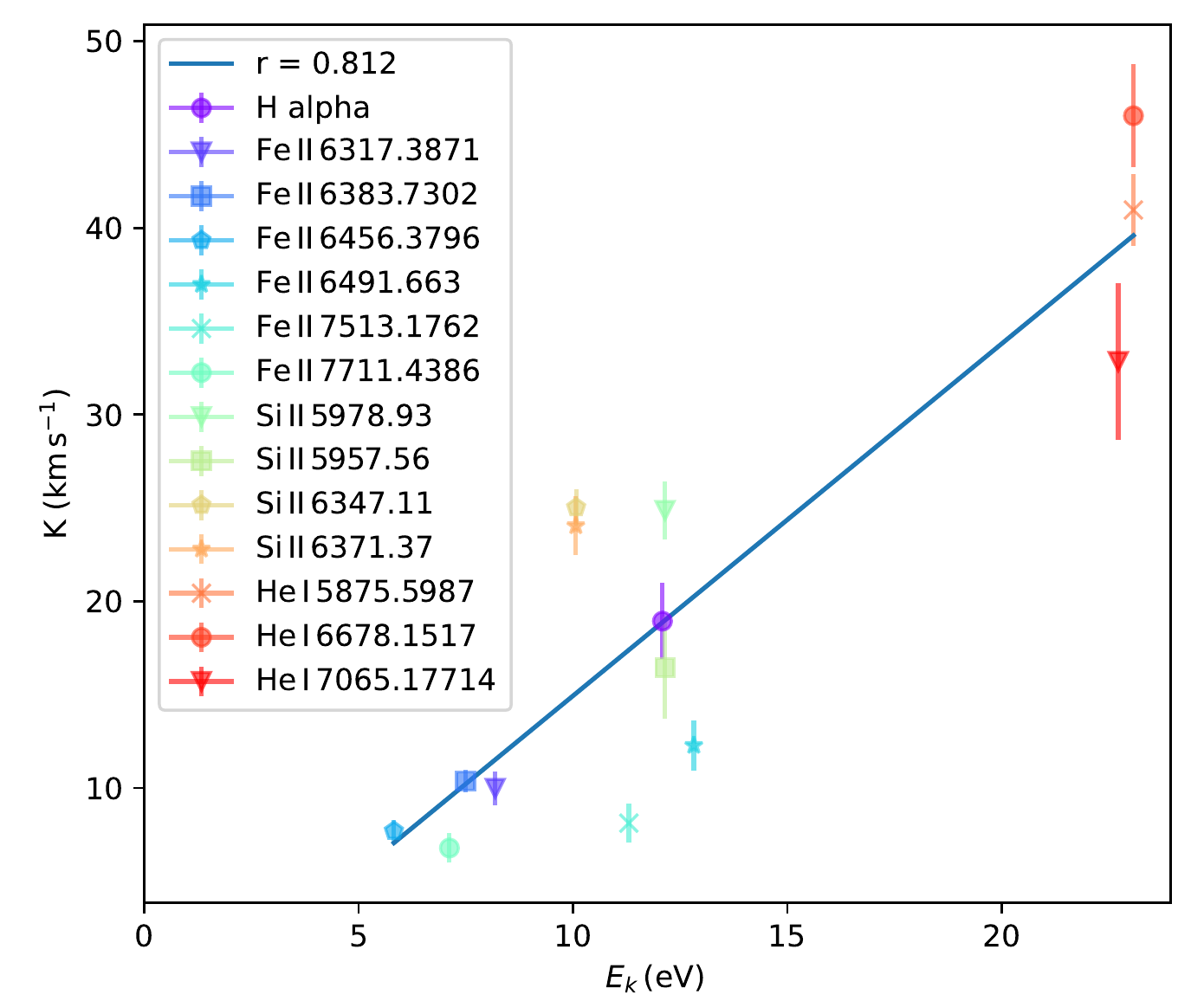}
    \caption{Amplitude, $K$, of the RV variations of the emission lines at the orbital period against $E_k$, the energy of the initial excited state leading to the lines. The correlation coefficient, $r$, weighted by the inverse of the square of the error, is quoted in the legend.}
    \label{fig:k_v_ek}
\end{figure}

As can be seen in Figures \ref{fig:he_31_rv_variability}, \ref{fig:si_31_rv_variability} and \ref{fig:fe_31_rv_variability}, the amplitudes and profiles of the RV variations are highly variable across different line species. The He\,I lines display the largest RV variations, the Si\,II emission then follows, and the Fe\,II lines display the smallest RV variations. Since the amplitudes of variations appear to be related to the species of emission line, this raises questions as to the origin of these variations. They cannot simply reflect the Keplerian motion of one of the binary components. To investigate, Figure \ref{fig:k_v_ek} displays $K$, the amplitude of the RV fits, against $E_k$, the energy of the upper atomic state of the transition associated with the emission line. Figure \ref{fig:k_v_ek} shows a strong positive correlation between $E_k$ and $K$: \textcolor{black}{the higher energy emission lines have larger observed changes in their radial velocities}. \\

\textcolor{black}{Since the emission lines also have differing $v_0$ by species, we investigate whether these are also correlated to $E_k$ in Appendix \ref{sec:v0_v_ek}. We find that there is no correlation between these parameters when all lines are included; however, if the Si\,II lines are excluded from the analysis there is a very strong correlation with higher $E_k$ lines having a more redshifted $v_0$. However, we cannot distinguish between whether the trend is real, with the Si\,II being anomalous, or whether there is no correlation between $v_0$ and $E_k$. Appendix \ref{sec:v0_v_ek} also investigates the relationship between the fitted jitter and $E_k$, finding that there is a strong correlation between these parameters. This indicates that lines with higher $E_k$ display more scatter in their RV variations at the orbital period.}\\

\section{Determining the orbital solution of GG Car using wind emission lines}
\label{sec:determining_orbital_solution}
\textcolor{black}{The orbital solution of the binary in GG Car has been as-of-yet inadequately determined. \cite{Marchiano2012} ascribed He\,I absorption as being due to photospheric absorption and they find a systemic velocity $v_0 = -162$\,\kms\ using this assumption. However, adopting this $v_0$ would mean that all emission lines in the system are significantly redshifted, implying that this $v_0$ is too negative (see the systemic velocity column of Table \ref{tab:orbital_solutions} and Figure \ref{fig:v0_v_ek}). Our inspection of FEROS and GJW spectra lead us to agree with \cite{Maravelias2018} that $v_0\approx -22$\,km\,s$^{-1}$ is more consistent with the radial velocities of the system's emission lines. Furthermore, the observation of photospheric absorption lines is rare in \sgBeshorthand s, and even in cases where they are observable they tend to be heavily polluted \citep{Kraus2016, Kraus2019ASupergiants}. The He\,I absorption being blueshifted absorption components originating in the strong stellar wind of the system is more consistent with the data, the \sgBeshorthand\ paradigm of opaque stellar atmospheres, the findings of \cite{Maravelias2018}, and \cite{Hanes2018StellarNebula}'s measurement of the average radial velocities of the stars in the Carina nebula. Therefore, a correct determination of the orbital parameters necessitates an alternative method to uncover the underlying Keplerian motion of the binary components.} \\

The emission lines we have studied that express RV variability at the orbital period of the binary must be formed in the circumstellar environment and therefore must be formed one of two ways, either in the circumstellar disk of one of the binary components, or in the stellar wind originating from one of the components. We can ascribe them to originating in the stellar wind rather than in a circumstellar disk for three reasons: the amplitudes of the RV variations are different for each line species, as shown in Figure \ref{fig:k_v_ek}, whereas they would be identical if they were formed in a disk which followed the Keplerian velocity of one of the stellar components; the primary, which would host the circumstellar disk, is almost filling its Roche lobe (\citealt{Kraus2013}; Section \ref{sec:enhanced_mass_transfer}, this study), making a circumstellar disk unstable; and the profiles of the lines which exhibit RV variability do not resemble the double-peaked profiles of Keplerian disks, as shown in the high-resolution FEROS spectra in Figures \ref{fig:feros_si_lines} and \ref{fig:feros_fe_lines}. The Fe\,II 6383.7 line even hosts a slight P-Cygni profile, with blueshifted absorption. The lines which do display disk profiles have no RV variability, indicating their formation in the circumbinary disk (Fe\,II 5991 and 6432, Figure \ref{fig:feros_fe_disk_lines}).\\

\begin{figure} 
  \centering
    \includegraphics[width=0.5\textwidth]{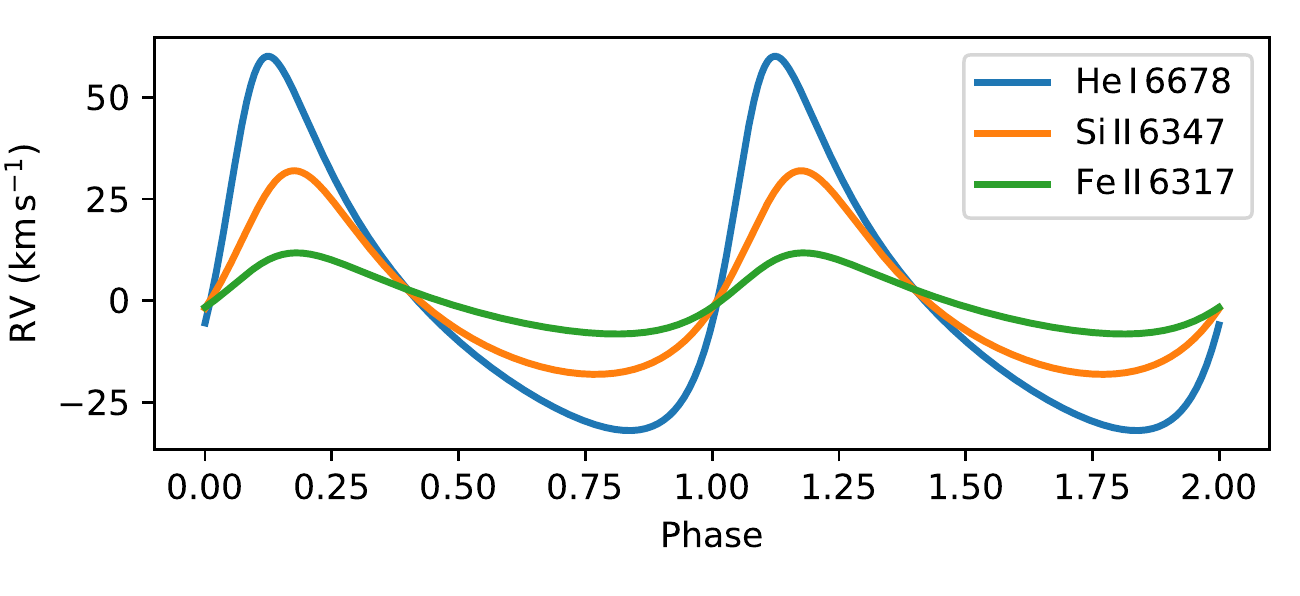}
    \caption{Comparison of the fitted RV models at the orbital period to representative He\,I, Si\,II, and Fe\,II emission lines. The systemic velocity has been subtracted from the models to highlight the synchronicity, but varying amplitudes, of the RV variations between the species.}
    \label{fig:rv_fits_comparison}
\end{figure}

The interpretation of the RV variations of the emission lines at the orbital period as originating in the stellar wind of the \sgBeshorthand\ primary explains the relationship of $K$ with $E_k$, as is shown in Figure \ref{fig:k_v_ek}. As is investigated in \cite{Grant2020Uncovering140} using the example of the Hydrogen Balmer series in $\eta$ Carinae, the measured velocities and therefore Keplerian orbital solutions from wind emission lines are dependent on the energetics of the emission lines (see their figure 8). Lines which have a higher $E_k$ are formed deep in the wind, closer to the star; conversely the lower $E_k$ lines are formed at larger radii and at later flow times on average than the higher energy lines. The lines are then formed for a finite time before the wind cools down below the temperature required to excite to the line's energy level. This smearing of the Keplerian velocity of the binary component leads to the higher energy lines tracing the Keplerian velocity of the star more closely than the lower energy lines. This smearing is demonstrated in Figure \ref{fig:rv_fits_comparison}: the He\,I, Si\,II and Fe\,II RV variations are largely in phase, but the lower energy lines have their variations smeared out compared to the highest energy He\,I line. \\

\cite{Grant2020Uncovering140}, using the examples of the luminous blue variable in $\eta$ Carinae and the Wolf-Rayet in RMC\,140, showed that the underlying orbital parameters of massive stars obscured in thick circumstellar envelopes can be probed by observing the radial velocity variations of the emission components of lines formed in the stellar wind, despite the smearing. They show that the underlying orbital motion can be found by fitting a Keplerian orbital solution which has been convolved with line emissivity kernels, which are calculated using stellar atmosphere codes to simulate the stellar winds. Therefore, we follow the methodology of \cite{Grant2020Uncovering140} to properly model the time-dependent emissivity of the individual wind lines in GG Car, and use these to fit the orbital parameters of the binary orbit. \\

\subsection{Methods}
\label{sec:methods}
The method of determining the Keplerian orbital solution using wind emission lines and line-formation kernels is described in exquisite detail in \cite{Grant2020Uncovering140}; here we outline the method in the context of determining the orbital solution of GG Car. \\

\subsubsection{CMFGEN simulations}
\label{sec:cmfgen}
The model of \cite{Grant2020Uncovering140} encapsulates both the star's orbital motion and the propagation of the wind, to reconcile differences in emission line variability with a consistent orbital solution. The model requires knowledge of the luminosity of each emission line as a function of wind-flow time outwards from the star, known as the line-formation kernels. \\

In order to calculate the line-formation kernels in GG Car emission, we simulate the atmosphere of the primary of GG Car using the non-local thermodynamic equilibrium, fully line-blanketed stellar atmosphere code, \texttt{CMFGEN} \citep{Hillier1998TheOutflows, Hillier2001CMFGEN:Winds}. \texttt{CMFGEN} allows us to calculate the luminosity of emission lines as a function of radius, $r$; we then translate this into wind-flow time, $t_{\rm flow}$, by recasting the radius using the analytic $\beta$-velocity law of \cite{Castor1975Radiation-drivenStars}:
\begin{equation}
    v_{\rm wind}(r) = v_0 + (v_\infty - v_0)\,\left(1-\frac{R_*}{r}\right)^\beta,
\end{equation}

\noindent where $v_\infty$ is the terminal velocity of the wind, $v_0$ is the initial velocity, and $R_*$ is the stellar radius, and integrating from the sonic point $r_s$ to $r$:
\begin{equation}
    t_{\rm flow} = \int_{r_s}^r \frac{dr}{v_{\rm wind}(r)}.
\end{equation} \\

Table \ref{tab:cmfgen_parameters} displays the important parameters used in the \texttt{CMFGEN} simulation of the atmosphere of GG Car. The stellar mass, luminosity, and radius are taken from Section \ref{sec:gaia_luminosity}, and the effective temperature is taken from the SED analysis of \cite{Marchiano2012}. The value of the terminal velocity, $v_\infty = 265$\,km\,s$^{-1}$, was chosen after inspection of the high-resolution FEROS spectra which show it is both the average half-width at zero intensity of the Si\,II emission, He\,I emission and H\,I Balmer emission from $n$=3--8, and the radial velocity of the blue-most edge of the middle-absorption component in the absorption complexes of the H\,I Balmer and He\,I lines. Appendix \ref{sec:v_inf} describes our determination of $v_{\infty}$ in further detail. The mass-loss rate, $\dot{M} = 2.2 \times 10^{-6}\,M_\odot\,{\rm yr}^{-1}$, was estimated using the wind momentum-luminosity relation for early B\,I-type supergiant stars \citep{Kudritzki2000WindsStars}. Our mass loss rate is similar to the measurement of \cite{Mcgregor1988ATOMICSTARS}.\\

\begin{table}
\centering          
\begin{tabular}{ l l}
\hline \hline
\\
   $M_{\rm *}$ & $24\,M_\odot$ \\ 
   $T_{\rm eff}$ & $23\,000$\,K \\
   $L_{\rm *}$  & $1.8\times10^5\,L_\odot$ \\
   $R_{\rm *}$ & $27\,R_\odot$ \\
   $\dot{M}$ & $2.2\times 10^{-6}\,M_\odot \, {\rm yr}^{-1}$\\
   $v_\infty$ & 265\,km\,s$^{-1}$ \\
   $\beta$ & 1 \\
   $f_{\rm v}$ & 0.1 \\
   $v_{\rm cl}$ & 100\,km\,s$^{-1}$\\
\end{tabular}
\caption{Parameters used in our \texttt{CMFGEN} simulations of the atomsphere of the \sgBeshorthand\ primary in GG Car. $M_{\rm *}$ is the mass of the star, $T_{\rm eff}$ is the effective temperature, $L_{\rm *}$ is the luminosity, $R_{\rm *}$ is the photospheric radius, $\dot{M}$ is the mass-loss rate, $v_\infty$ is the terminal velocity, $\beta$ is the velocity law exponent, $f_{\rm v}$ is the clumping volume filling factor, and $v_{\rm cl}$ is the wind velocity where clumping becomes important.} 
\label{tab:cmfgen_parameters}      
\end{table}

We choose $\beta$ to be 1, as this has been shown to be a typical value for supergiants in dense winds \citep{Puls1996O-starPredictions.}. The clumping parameters of the filling factor, $f_{\rm v}$, and the representative wind clumping velocity, $v_{\rm cl}$, having no previous determination in the literature on GG Car and being indeterminate from the few high-resolution FEROS spectra available, were chosen to be standard values for massive stars at 0.1 and 100\,km\,s$^{-1}$ respectively. \\

\begin{figure}
  \centering
    \includegraphics[width=0.5\textwidth]{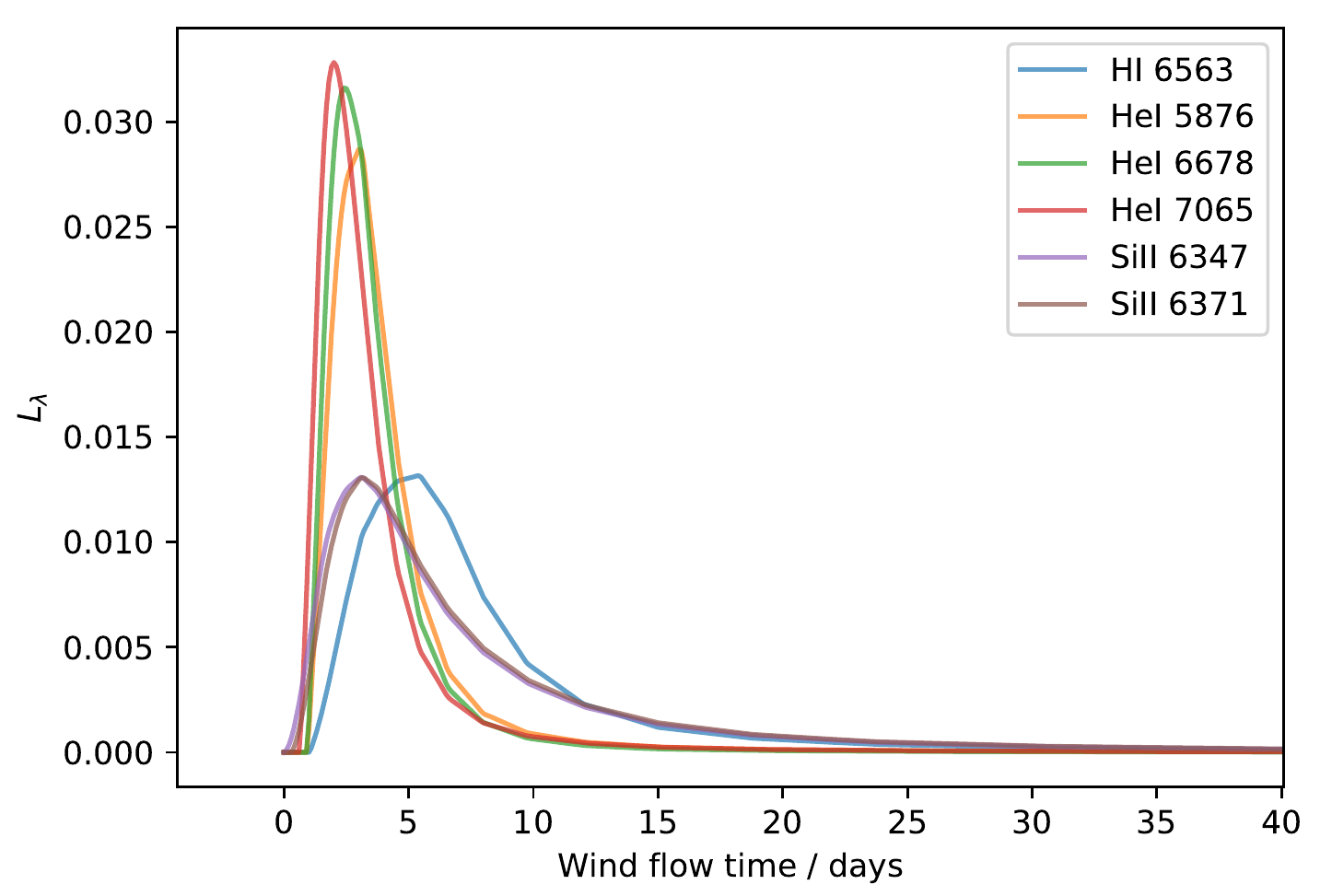}
    \caption{Emission line emissivity, $L_\lambda$, against wind-flow time for the wind emission lines used in this study.}
    \label{fig:line_formation_kernels}
\end{figure}

The resulting line luminosities as a function of wind-flow time are shown in Figure \ref{fig:line_formation_kernels}. The calculated line luminosities against wind-flow time shown in Figure \ref{fig:line_formation_kernels} reflect that the He\,I lines are formed shortly after the wind passes the sonic point, whereas the Si\,II and H-alpha emission form later on average. Table \ref{tab:line_kernel_times} lists the peak and median emission times of the calculated line luminosities. The peak and median times of the line luminosities for all lines are less than the $\sim$31-day period of the binary. We were unable to extract line-emissivity kernels of Si\,II 5957 and 5978 or any Fe\,II lines from the \texttt{CMFGEN} simulation, so these lines are excluded from this analysis. \\

\begin{table}
\centering          
\begin{tabular}{ l l l}
Line & Peak [days] & Median [days] \\
\hline 
H-alpha & 5.4 & $6.4^{+6.0}_{-2.7}$  \\
He\,I 5876 & 3.0 & $3.5^{+2.3}_{-1.2}$  \\
He\,I 6678 & 2.4 & $3.2^{+2.0}_{-1.1}$  \\
He\,I 7065 & 2.0 & $2.9^{+2.3}_{-1.1}$  \\
Si\,II 6347 & 3.1 & $5.5^{+8.5}_{-3.0}$  \\
Si\,II 6371 & 3.1 & $5.8^{+8.8}_{-3.0}$  \\
\end{tabular}
\caption{Time to the peak and median emission of the lines' emissivity kernels, calculated with \texttt{CMFGEN}. The difference of the median time with the 84th and 16th percentiles of the emission are quoted by the superscript and subscript respectively.} 
\label{tab:line_kernel_times}      
\end{table}

\subsubsection{Fitting the orbital solution}

Section \ref{sec:spectroscopic_variability} describes how we fit orbital solutions to the individual emission lines' fits; from these fits we extract the jitter, $j$, and systemic velocity, $v_0$, for each individual line. $j$ and $v_0$ for each line are listed within Table \ref{tab:orbital_solutions}.\\

We then fit one Keplerian orbital solution to the RV data of all emission lines simultaneously using the line emissivity kernels described in Section \ref{sec:cmfgen}. We fit the orbital solution by maximising the log-likelihood function, again using the MCMC algorithm \texttt{emcee}. In the first instance, we subtract the $v_0$ calculated in the single line-fitting routines from the data for each emission line. Then, for each MCMC iteration, we convolve the Keplerian solution, $v_{\rm kep}$, for the given input parameters, $\theta$, with the line formation kernel, $\lambda_k$, of each line and calculate the log-likelihood using the convolved Keplerian velocities. The log-likelihood function for the convolution model is thereby given as
\begin{multline}
\begin{split}
    \text{ln} \, P (\theta \mid D, \, \sigma) = - \frac{1}{2} \sum_{k}\sum_{i=0}^N \left[\frac{\left(D_{k,i} - \lambda_k*v_{\rm kep}(\theta) - v_0\right)^2}{\sigma_{k,i}^2 + j_k^2}  \right.\\ \left.+ \text{ln}\left(2\pi (\sigma_{k,i}^2 + j_k^2)\right)\right],
\end{split}
\end{multline}
\noindent where $D_{k,i}$ is the $i$th data point for line $k$, $\sigma_{i,k}$ is the $i$th uncertainty for line $k$ taken from the Gaussian fitting routine, $\lambda_k$ is the line formation kernel for line $k$, and $j_k$ is the jitter for line $k$. $j_k$ for each line is taken from the single line fitting routine, described in Section \ref{sec:spectroscopic_variability}, to account for scatter in the data. \\

Therefore, in this method, we are fitting for one unified set of the primary's orbital parameters, $\theta$, encoding its semi-amplitude, $K$, argument of periastron, $\omega$, eccentricity, $e$, phase of periastron, $M_0$, and the orbital period, $P$, using all emission lines simultaneously. $M_0$ is the phase of periastron relative to a reference time $T_0$; we chose $T_0$ to be JD\,2452051.93, which is the time of periastron found by \cite{Marchiano2012}. We also fit for $v_0$ to aid the fitting routine, though this will not be the true systemic velocity of the system, but instead should be near zero since the data for each line has had its individual systemic velocity subtracted.\\

We run 256 MCMC walkers for 10\,000 burn iterations then 10\,000 walk iterations. The fitted parameters are taken as the median of the Monte Carlo samples of the parameters, and the errors the difference of the median with the 16th and 84th percentiles.\\ 

\subsection{Orbital solution fit results}
\label{sec:orbital_solution_results}
\begin{table}
\centering          
\begin{tabular}{ l  l}
\hline
\hline \\

$K$	& $48.57^{+2.04}_{-1.87}$\,\kms\\
$\omega$	& $339.87^{+3.10}_{-3.06}$\,$^\circ$	\\
$e$	& $0.50^{+0.03}_{-0.03}$	\\
$v_0$	& $-0.72^{+0.36}_{-0.36}$\,\kms	\\
$M_0$	& $202.35^{+15.16}_{-15.01}$\,$^\circ$ \\
$P$ & $31.01^{+0.01}_{-0.01}$\,days	\\
$T_{\rm peri}$ & ${\rm JD}\,2452069.36\pm1.30$ \\
\end{tabular}
\caption{The fitted values for the underlying Keplerian motion of the \sgBeshorthand\ primary using all emission lines in aggregate and their line formation kernels, $\lambda_k$. $K$ is the orbit amplitude, $\omega$ is the argument of periastron, $e$ is orbital eccentricity, $v_0$ is the velocity offset without physical significance, $M_0$ is the phase offset of the reference time $T_0 = {\rm JD}\,2452051.93$, $P$ is the orbital period, and $T_{\rm peri}$ is the epoch of periastron.} 
\label{tab:fit_parameters}      
\end{table}

Table \ref{tab:fit_parameters} presents the results for the underlying Keplerian orbital solution, using all emission lines simultaneously, and Figure \ref{fig:convolved_fitted_keplerian_corner} in the Appendix displays the posterior probability distribution for the fitted orbital parameters. We find that the binary has a period of $31.011\pm0.01$\,days, and is very eccentric. The phase of periastron $M_0$, compared to the reference time $T_0={\rm JD}\,2452051.93$ is $202.35^\circ$, meaning that we find the ``phase'' of periastron is significantly different to \cite{Marchiano2012}. We find time of periastron $T_{\rm peri} = {\rm JD}\,2452069.36\pm1.30$, which therefore gives us the binary orbital ephemeris of Equation \ref{eq:ephemeris}.\\

\begin{figure}
  \centering
    \includegraphics[width=0.5\textwidth]{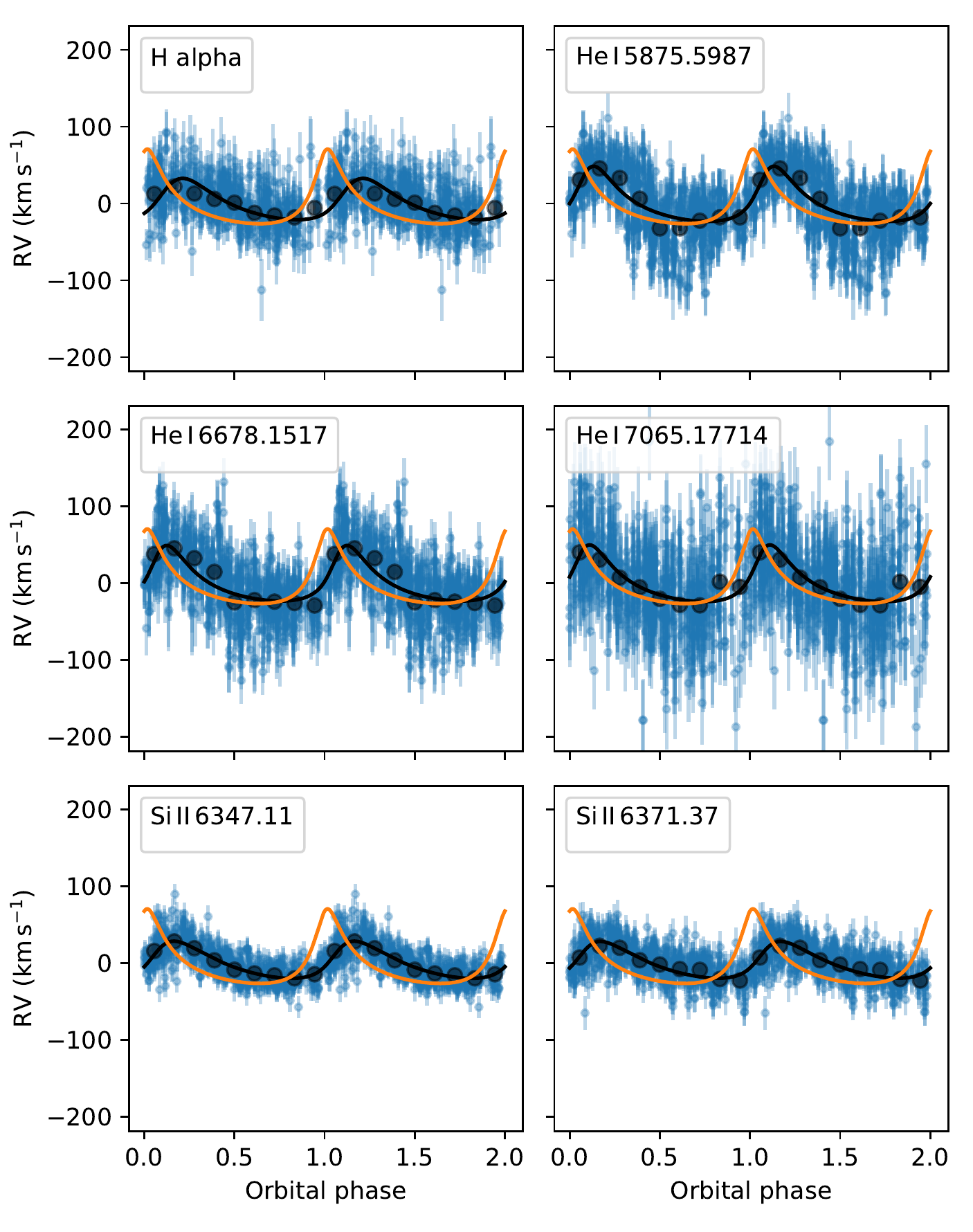}
    \caption{Results of the fitted Keplerian solution convolved with the line formation kernel $\lambda_k$ for each line. The underlying Keplerian velocity, which is the same for each emission line's panel, is plotted in orange. The Keplerian velocity convolved with the line formation kernel is plotted as a black line. Black points show the average RV of the data in 10 phase bins. The data for each line has had its systemic velocity listed in Table \ref{tab:fit_parameters} subtracted.}
    \label{fig:convolved_fitted_keplerian}
\end{figure}

Figure \ref{fig:convolved_fitted_keplerian} displays the data for the RV of each emission line in this study, along with the fitted underlying Keplerian motion in orange, and the Keplerian motion convolved with the line's formation kernel, $\lambda_k$, as a black line. Black points denote the average value of the RV data in 10 phase bins. The convolved Keplerian velocities match the phase averaged RV data exceptionally well for all emission lines studied, indicating that this orbital solution along with the calculated $\lambda_k$s for each line can recreate the measured RV variability. \\

\textcolor{black}{Therefore, though \texttt{CMFGEN} assumes spherical symmetry and the winds of \sgBeshorthand s are inherently non-symmetric, we find that simulating the mid-velocity wind speed gives line-formation kernels which are consistent with the relative RV variations of the emission lines in GG Car. The present study, therefore, marks a significant improvement in the determination of the orbital solution of the binary. Further studies may simulate the line formation kernels expected in an inhomogeneous wind, in order to extract an even more accurate orbital solution.} \\

\section{Discussion}
\label{sec:discussion}

\subsection{Constraints on the mass of the secondary}
\label{sec:secondary_mass}
Since we do not detect any spectral lines from the secondary, we can not make direct inferences about its mass; however, we can place constraints on its mass using the orbital solution of the primary. The binary mass function for a single-lined binary, $f$, is given by
\begin{equation}
\label{eq:mass_function}
    f = \frac{M_2^3\,\sin^3{i}}{(M_1 + M_2)^2} = \frac{PK^3}{2\pi G}(1-e^2)^{3/2}
\end{equation}
\noindent where $M_1$ is the mass of the binary component whose radial velocities have been measured, $M_2$ is the mass of the unseen companion, $i$ is the inclination of the orbit, $P$ is the orbital period, $K$ is the amplitude of the binary orbit, $e$ the eccentricity of the orbit, and $G$ the gravitational constant. $f$ gives the lower limit of $M_2$ as
\begin{equation}
    M_2 > \text{max}(f, f^{1/3}M_1^{2/3}).
\end{equation}

\noindent \textcolor{black}{Using the new orbital parameters in Table \ref{tab:fit_parameters} gives a mass function of $f=0.24\pm0.03\,M_\odot$. Using the primary mass, $M_1 = M_{\rm pr} = 24 \pm 4 M_\odot$, we can infer that $M_2>5.2\pm 0.6 \, M_\odot$.} \\

However, there is also the indication from \cite{BorgesFernandes2010THEEYES} and \cite{Kraus2013} that the circumbinary disk in GG Car has an inclination of $\sim$\,$60 ^\circ$ to the line of sight. If we were to assume that the orbit of the binary and circumbinary disk are roughly co-planar, then we can use the circumbinary disk inclination as an estimate for the inclination of the binary orbit. Therefore, we adopt a value of $i = 60 \pm 20 ^\circ$, allowing a large uncertainty to acknowledge both the uncertainty from these studies and any precession of the circumbinary disk in relation to the binary orbit \citep{Doolin2011}. With an estimate for the binary inclination, we can solve Equation \ref{eq:mass_function} for $M_2$. \textcolor{black}{This gives $M_2 = 7.2^{+3.0}_{-1.3}\,M_\odot$, using Monte Carlo means to calculate the uncertainties. For this secondary mass, we can calculate the semi-major axis of the orbit as $a=0.61\pm0.03$\,AU. This makes the mass ratio $q=M_{1}/M_{\rm 2} = 3.3^{+0.6}_{-1.0}$.}\\

\textcolor{black}{Here we question whether the secondary of GG Car could significantly contribute to the V-band brightness of the system, and therefore affect the determination of the luminosity of the primary in Section \ref{sec:gaia_luminosity}. If the binary components have evolved as pseudo-single stars, we can infer that the secondary is still on the main sequence since the primary is expected to have only recently evolved off the main sequence \citep{Kraus2009, Kraus2013}. With its mass, the luminosity of the secondary would be $\sim$700\,--\,4800\,$L_{\odot}$, which is between 0.4--2.7\% of the luminosity of the primary. In the case that there has been binary interaction in GG Car's past, \cite{Farrell2019ImpactSupergiants} finds that the secondary components of blue supergiants in binaries should have negligible contributions on the V-band photometry, except in cases where the mass ratio is near unity. In a scenario of full mass inversion, where the current secondary is a stripped object, the stripped companion would be expected to be less luminous than the primary and would also emit most strongly in the UV \citep{Wellstein2001FormationBinaries}. Even in extreme cases of very bright stripped companions, they are expected to be outshone at visible wavelengths by their primaries \citep{Gotberg2018SpectralStars}. We can therefore conclude that, in all evolutionary cases, the secondary's contribution to the V-band brightness of GG Car is negligible and the luminosity determination of the primary in Section \ref{sec:gaia_luminosity} remains valid.}\\

\subsection{Photometric variations along the binary orbit}
\label{sec:photometric_variations}
The photometric variation of GG Car along the orbit of the binary is continuous with one minimum and one maximum. The average peak-to-trough magnitude change over the binary orbit is around 0.2\,mag in the V-band, corresponding to a brightness contrast of around 20\%. \textcolor{black}{In the literature, GG Car has occasionally been referred to as an eclipsing binary; however, there are no signs of eclipses in the lightcurve which survive the scatter in the data and, furthermore, eclipses are not expected to occur for our measured semi-major axis of the orbit, primary radius, and the reported orbital inclination.} \\

\begin{figure}
  \centering
    \includegraphics[width=0.5\textwidth]{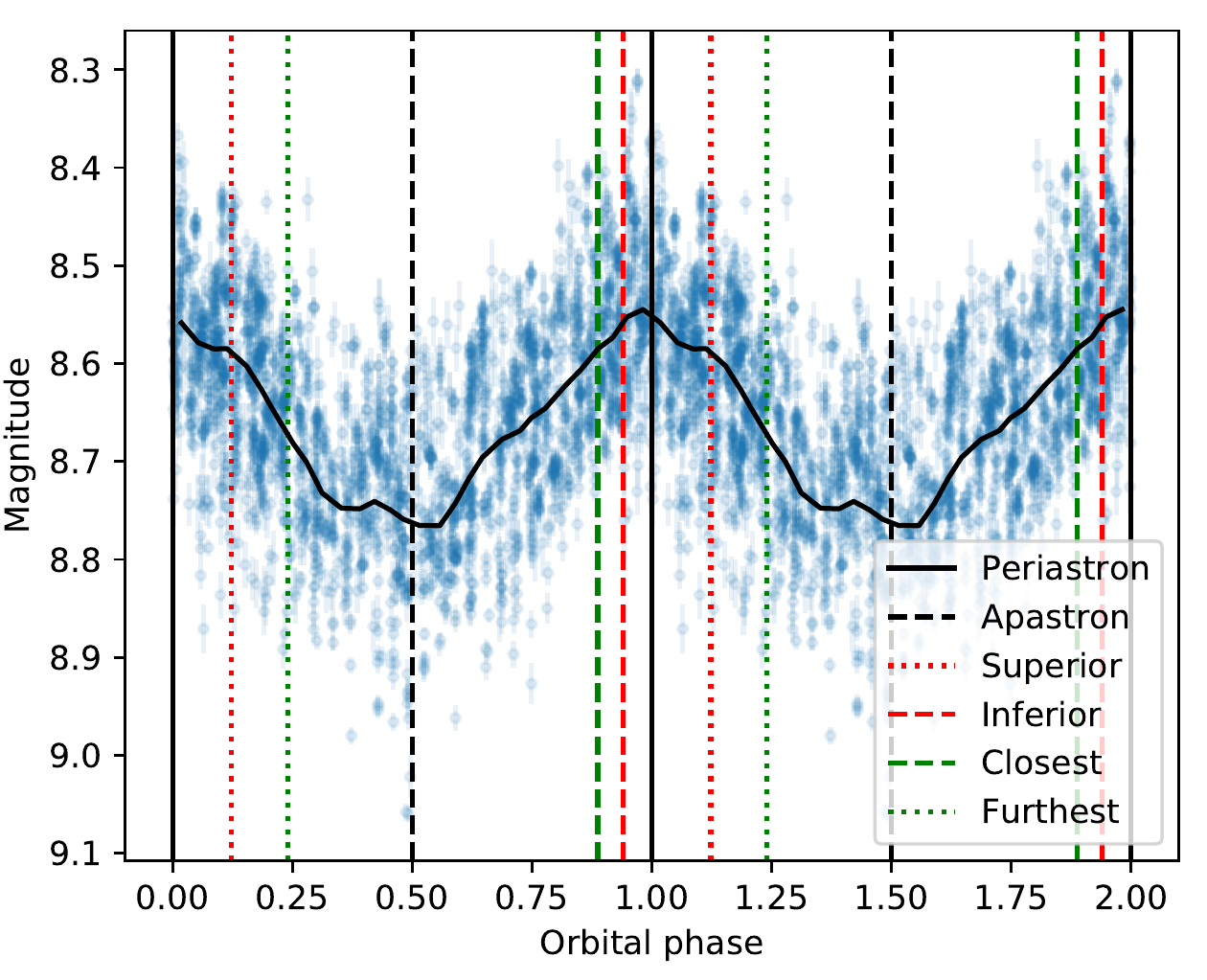}
    \caption{The same as Figure \ref{fig:v_band_31_fold}, except important phases in the  Black vertical line indicates phases of periastron and apastron, vertical red lines phases of superior and inferior conjunction of the \sgBeshorthand\ primary, and green lines indicate when the \sgBeshorthand\ primary is closest or furthest from the observer in its orbit. The Keplerian radial velocity of the \sgBeshorthand\ primary is overplotted as a gray line, on the right hand axes.}
    \label{fig:v_band_31_fold_with_orbit}
\end{figure}

Figure \ref{fig:v_band_31_fold_with_orbit} shows how the photometric variations correspond to the orbital solution determined in Section \ref{sec:determining_orbital_solution}, with pertinent stages of the binary orbit noted. The phase at which the system is brightest corresponds near exactly with periastron, peak brightness occurring just slightly before, and correspondingly the system is dimmest at apastron. There is no perceptible dimming at either superior or inferior conjunction, further showing that the binary is not eclipsing. Figure \ref{fig:v_band_31_fold_with_orbit} also marks phases where the primary is closest and furthest from the observer in its orbit. \\

In the following sections we explore some causes of continuous photometric variations at the orbital period. In Section \ref{sec:reflection} we rule out reflection between the binary components and in Section \ref{sec:extinction} we rule out extinction or reflection by circumbinary material as causes of the brightness changes observed in GG Car. In Section \ref{sec:enhanced_mass_transfer}, we show that enhanced mass transfer between the binary components and accretion at periastron is very likely to occur given the orbital parameters, and that these can reproduce the photometric variability observed assuming a simple model.  \\

\subsubsection{Reflection}
\label{sec:reflection}

A possible explanation of the peculiar lightcurve of GG Car is reprocessing and reemission of the radiation incident on each stellar component \citep{Wilson1990AccuracyEffect}. If the variations were dominated by the reflection of one component's radiation by its companion, the system's lightcurve would have one maximum and one minimum per orbital period as the hemisphere of the companion which is illuminated orbits in and out of the line of sight. \textcolor{black}{As is shown in Figure \ref{fig:v_band_31_fold_with_orbit}, the photometric maximum occurs near inferior conjunction, when the primary and the secondary line-up along the line-of-sight, with the primary being closer to the observer than the secondary.}\\

Entertaining a scenario where the primary heats the secondary, or a disk around the secondary, and the luminosity intercepted is re-radiated we can presume, in the best case scenario, that the fractional change in luminosity would be given by 
\begin{equation}
\frac{\Delta L}{L} \approx \frac{\Delta L}{L_{\rm pr}} = \frac{ \pi R_{\rm sec}^2}{4\pi ((1-e)a)^2},
\end{equation}
\noindent where $R_{\rm sec}$ is the radius of the secondary, $e$ is the eccentricity of the orbit, $a$ is the semi-major axis of the orbit, $L$ is the total luminosity of the primary and the secondary, and $L_{\rm pr}$ is the luminosity of the primary. In this best case scenario, we approximate the intrinsic luminosity of the secondary, $L_{\rm sec} \approx 0$. Since the brightness contrast is $\sim$20\%, in order for reflection to account for the photometric variation, with $e=0.5$ and $a=0.61$\,AU, $R_{\rm sec}$ would have to be 0.27\,AU. Using the approximation \cite{Eggleton1983ApproximationsLobes}, the radius of the secondary's Roche lobe at periastron can be calculated within 1\% accuracy by
\begin{equation}
\label{eq:roche}
    r_{\rm sec}= \frac{0.49 q^{2/3}}{0.6q^{2/3} + \ln \left(1 + q^{1/3}\right)}\,(1-e)\,a,
\end{equation}

\noindent where $q$ is the mass ratio $M_{\rm sec}/M_{\rm pr}$. Given a mass ratio of 7.2/24, the size of the secondary's Roche lobe at periastron is 0.09\,AU; therefore, the secondary cannot both be dynamically stable and have a large enough size to reflect enough of the primary's light in order to account for the observed variability. If the secondary filled its Roche lobe, it would lead to a maximum luminosity contrast of $\sim$2.0\%; therefore, reflection of the primary's light off the secondary would only provide an insignificant amount of the observed flux change. This is also a best case scenario, which assumes that at maximum light the full illuminated hemisphere of the secondary is facing the observer and at minimum light the illuminated hemisphere is facing away from the observer; given that the system is at $\sim$60$^\circ$ inclination and its argument of periastron is almost in the plane of the sky, neither statement is true, and this would diminish the luminosity contrast due to reflection further.\\

Should the secondary's luminosity be non-negligible and heat the primary, we can estimate the luminosity of the secondary required to heat the primary in order to facilitate the observed contrast, $\Delta L / L = 0.2$, where $L = L_{\rm pr} + L_{\rm sec}$. We can express the luminosity of the secondary $L_{\rm sec}$ as 
\begin{equation}
    L_{\rm sec} = (0.2 L_{\rm pr})\ /\ \left(\frac{\pi R_{\rm pr}^2}{4 \pi ((1-e)a)^2}-0.2\right).
\end{equation}
\noindent Entering the orbital parameters and $R_{\rm pr}=27\,R_\odot$, $L_{\rm sec}$ would have to be negative in order to account for the change. This is clearly unphysical. \textcolor{black}{Additionally, the brightness of the system is falling around superior conjunction, which is incompatible with the brightness changes being caused by reflection of the secondary's flux by the primary.}\\

Therefore, given the stellar and binary orbital parameters of GG Car, we can conclude that reflection is not a significant source of the photometric variability over the orbital period. \\

\subsubsection{Extinction, scattering, and reflection by circumbinary material}
\label{sec:extinction}

Circumbinary orbiting material has been invoked to explain photometric variability of binaries at their orbital periods, attributing this to variable extinction, scattering and reflection of starlight by the circumbinary material (e.g. \citealt{Waelkens1991VariabilitySystem., Waelkens1996VariabilityNebula, Khokhlov2018, Ertel2019ResolvedDisk}). Given its thick, probably near opaque, circumbinary envelope, could similar processes be causing the photometric variability in GG Car? \\

Should scattering or reflection of the primary's light by the circumbinary disk be occuring, the brightest times of the system would be when the primary is closest to the far side of the circumbinary disk, which is when the primary is furthest from the observer. Figure \ref{fig:v_band_31_fold_with_orbit} notes at which point along the light curve that the primary is furthest from the observer, which is when the system is at roughly its mean brightness, indicating that variable scattering and reflection of the primary's light by the far edge of the circumbinary is most likely not the source of the photometric variations at the orbital period. Conversely, if there were considerable variation in the extinction of the primary's flux by the circumstellar envelope over the binary period (as was proposed in GG Car by \cite{Gosset1985}), then the system should be brightest when the primary is closest to the observer. Again, this is not observed, therefore we rule out these possibilities.\\

\subsubsection{Enhanced mass transfer and accretion at periastron}
\label{sec:enhanced_mass_transfer}
Given that the brightness of GG Car is maximal remarkably near to periastron, might the brightness variations be due to enhanced mass loss and mass transfer between the binary components at periastron? This behaviour is widely seen in ``Type I'' outbursts in Be/neutron star X-ray binaries (see \citealt{Reig2011Be/X-rayBinaries} for a review). At periastron, the neutron star travels through the decretion disk of the Be star, leading to an increase in luminosity.\\

\begin{figure}
  \centering
    \includegraphics[width=0.5\textwidth]{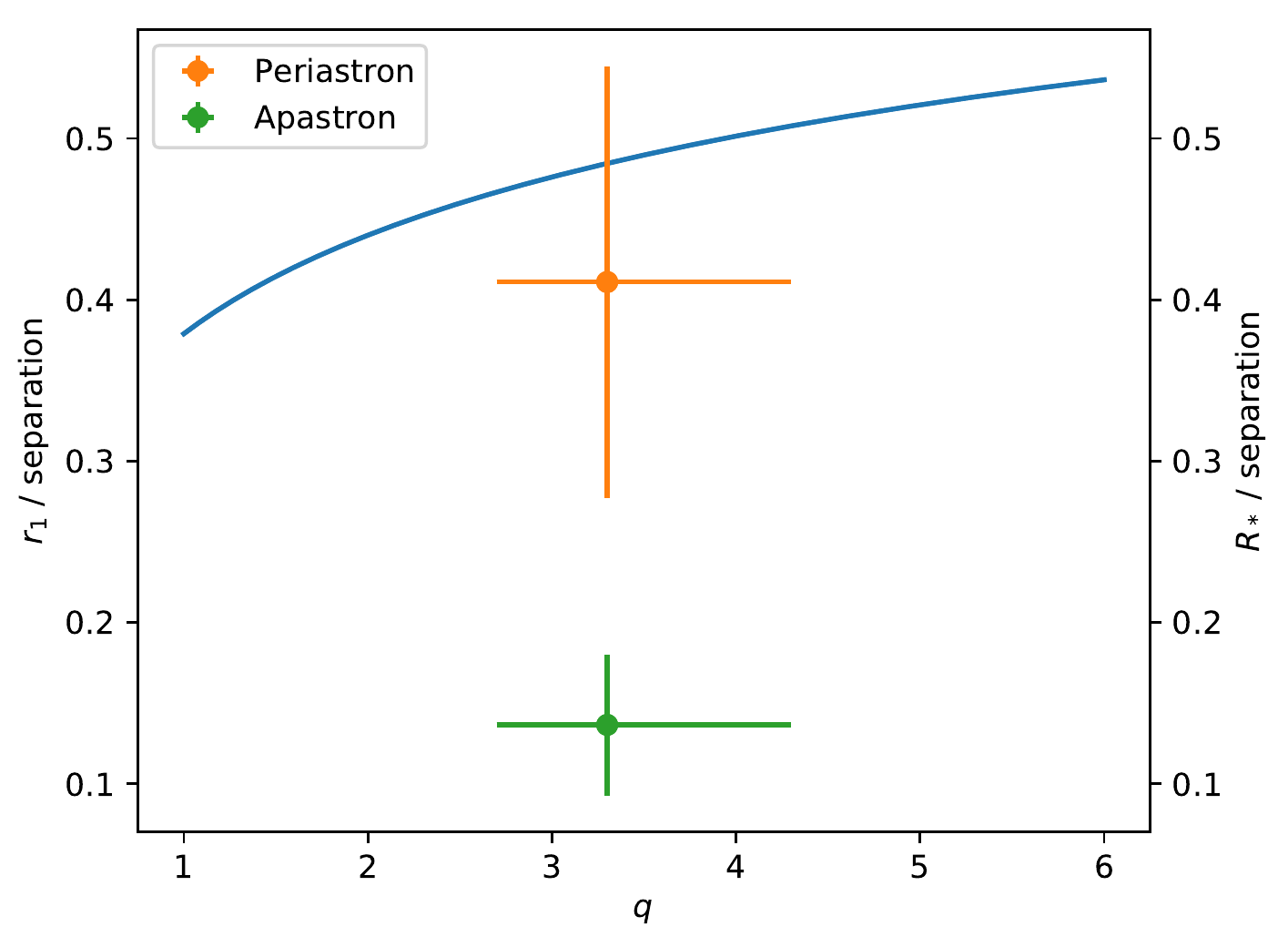}
    \caption{Size of the primary's Roche lobe as a function of binary mass ratio, $q$, in units of the binary separation (blue line), and the radius of the primary, $R_*$, in units of the binary separation at periastron (orange point) and apastron (green point).}
    \label{fig:primary_radius_v_roche_lobe}
\end{figure}

Figure \ref{fig:primary_radius_v_roche_lobe} demonstrates the relative sizes of the primary and its Roche lobe against mass ratio at periastron and apastron. With a binary mass ratio of $q = 3.3^{+0.6}_{-1.0}$, separation of $0.61\pm0.03$\,AU, eccentricity $0.50\pm0.03$ and $R=27\,R_\odot$, the primary radius extends to $85 \pm 28$\,\% of its Roche lobe's radius at periastron. Conversely, at apastron the primary radius extends to $28 \pm 9$\,\% of the Roche radius. This indicates that the wind of the primary would have to travel significantly further at apastron than periastron before it encounters the secondary. \textcolor{black}{\cite{Moreno2011EccentricBinaries}, calculating the rate of energy dissipation of tidal flows in the surfaces of stars in binaries, calculates that stellar mass-loss may be enhanced around periastron in eccentric orbits due to the increased rate of energy dissipation at periastron. They additionally show that this mass-loss would be expected to be focused in the equator of the star in the orbital plane, such as is expected in \sgBeshorthand s. The trailed spectra of the He\,I lines shown in Figure \ref{fig:he_trailed_spectra} show that the blueshifted absorption component becomes deeper around phase 0; this likely indicates that more mass is being lost around periastron in the primary's wind.} This all indicates that the rate of mass-loss of the primary and the rate of mass being captured by the secondary will be strongly dependent on the phase of the binary given the orbit's eccentricity, and therefore we can expect that there will be changes in the brightness due to varying rates of mass transfer and accretion in GG Car. Any rotation of the primary would additionally shrink its effective Roche lobe, aiding mass loss at the stellar equator. \\

\textcolor{black}{To investigate whether accretion of the primary's equatorial wind by the secondary may cause the observed V-band flux changes over the orbital period, we construct a two dimensional model of the binary with the primary's equatorial wind modelled as a series of equal-mass rings emitted every time-step from its equator with wind speed $v_{\rm wind}$, whose bulk velocity travels at the velocity of the binary at emission. We assume that $v_{\rm wind}$ is constant for this simple calculation, that the mass loss rate is also constant, and that the wind is confined to near the equator and is concentrated in the orbital plane. We count a ring as being accreted when it crosses into the Roche lobe of the secondary. When a ring is accreted in the simulation we set the amount of mass accreted from it to}
\begin{equation}
    \label{eq:accretion_weighting}
    \Delta M_{\rm ring} = \frac{K}{r_{\rm ring}\,v_{\rm diff}},
\end{equation}

\noindent where $r_{\rm ring}$ is the radius of the ring, $v_{\rm diff}$ is the differential speed between the secondary and the part of the ring being accreted, and $K$ is an unimportant constant. Once a part of a ring has been accreted, the rest of the material in the ring is assumed to carry on expanding unimpeded away from the binary and is ignored. In our model, we therefore calculate the mass accretion rate in a timestep
\begin{equation}
\label{eq:simulated_accretion_rate}
    \dot{m} = \sum_{i=0}^{N_{\rm acc}} \Delta M_{{\rm ring},\,i},
\end{equation}
\noindent where $i$ runs from 0 to $N_{\rm acc}$, the number of rings accreted in the timestep. \textcolor{black}{Appendix \ref{sec:accretion_model} describes this simple two-dimensional model, and the reasoning behind the weighting of Equation \ref{eq:accretion_weighting}, in more detail.}\\

\begin{figure}
  \centering
    \includegraphics[width=0.5\textwidth]{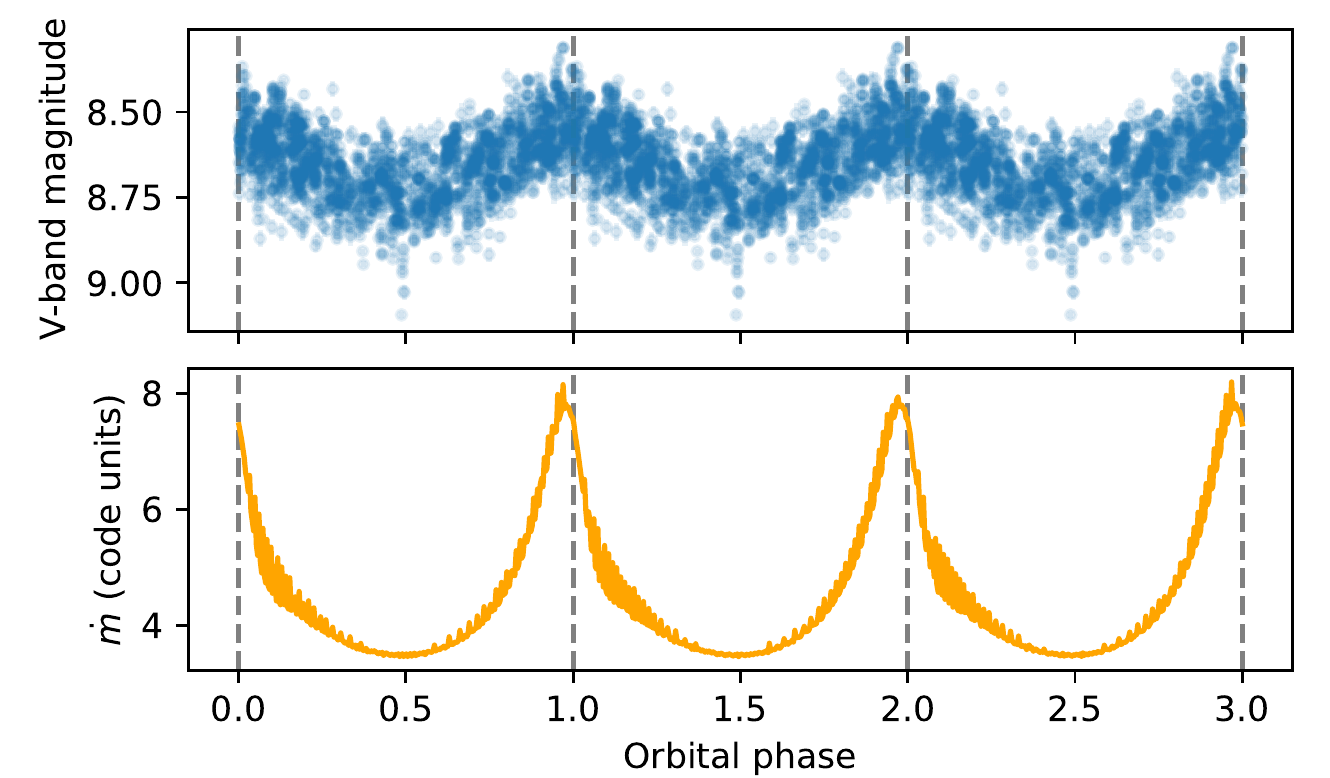}
    \caption{\textcolor{black}{Top panel: The V-band photometry of GG Car by orbital phase. Bottom panel: The simulated mass accretion rate of the secondary $\dot{m}$, in arbitrary code units. Grey dashed lines indicate phases of periastron. The simulated luminosity change is calculated using Equation \ref{eq:simulated_accretion_rate}. The very short term variability in the simulated $\dot{m}$ is numerical noise from the calculation. The simulated shape of the mass accretion rate is strongly correlated with the changes in the V-band photometry over the binary orbit.}}
    \label{fig:accretion_photometry}
\end{figure}

Using the GG Car orbital parameters of $e=0.50$, $a=0.61$\,AU, $M_{\rm pr}=24\,M_\odot$, $M_{\rm sec}=7.2\,M_{\odot}$, $R_*=27\,R_\odot$, $v_{\rm wind}=265$\,\kms, we calculate the expected changes in accretion rate of GG Car over four orbital periods. The results of our calculation compared with GG Car's V-band photometry are shown in Figure \ref{fig:accretion_photometry}. The simulation takes around one orbital period to reach a steady state, so we excise the first orbit from Figure \ref{fig:accretion_photometry} to avoid misleading edge effects. The simulation results in Figure \ref{fig:accretion_photometry} have not been folded by orbital phase, but the simulation has reached a steady state for each orbit.\\

\textcolor{black}{The shape of the simulated mass accretion rate in this two-dimensional model and the observed V-band photometry are remarkably correlated over the orbital period; the peaks of the expected accretion rate match the peak V-band photometry closely, with the peak brightness occurring just before periastron and minimum brightness at apastron. Appendix \ref{sec:accretion_model} investigates how changing the eccentricity and the wind speed affects the calculation results and finds that the resulting accretion rate variability is similar for higher wind speeds, but that the eccentricity should be between $\sim0.2$\,--\,0.7 to match the observed V-band variability. This, and the peak brightness occurring at periastron, lends supporting evidence for the orbital solution and the time of periastron passage found in Section \ref{sec:orbital_solution_results}, which found $e = 0.50\pm0.03$. A wind accretion model may also explain the stochastic variances of the lightcurve's profile over successive orbital cycles: since the luminosity changes will be heavily dependent on the conditions of the equatorial wind, any fluctuations or changes in the wind will affect how the brightness changes over an orbital period. }\\

We therefore propose that two-dimensional accretion is powering the V-band photometric variability of GG Car over its orbital period, with the secondary accreting the equatorial wind of the primary. \textcolor{black}{The increase in V-band luminosity in this scenario will be due to heating from the wind falling into the gravitational well of the secondary; however, whether this heating is of the secondary's surface or of an accretion disk around the secondary cannot be yet determined. Furthermore, any accretion luminosity which occurs in the secondary is likely to be reprocessed by the thick circumstellar envelope of the system. Concurrent photometric surveys in the UBV bands, along with high time- and spectral-resolution spectroscopy focused around periastron to search for accretion signatures, would help to determine the correctness of this scenario.}\\

\textcolor{black}{This calculation aims to provide a first-order, proof-of-concept calculation of the variable accretion rate of the primary's equatorial wind by the secondary through the eccentric orbit of the binary components. We make no attempt to fully quantify the absolute value of the accretion rates, nor the full calculation of the expected changes in magnitude in the V-band, which are beyond the scope of the present study. We are making the assumption that the rate of mass accretion by the secondary is proportional to the accretion luminosity in the V-band. In addition, we do not attempt to model time-varying mass-loss rates of the primary, which would be expected in the varying gravitational influence of the secondary during the eccentric orbit; however, this effect would be expected to increase the mass-loss rate, and therefore also increase the accretion rate of the secondary, at periastron \citep{Lajoie2011MASSMETHOD, Lajoie2011MASSBINARIES, Moreno2011EccentricBinaries, Davis2013MassBINSTAR}.} \\

\textcolor{black}{A scenario of mass-loss and mass-transfer focused at periastron could account for the large eccentricity of GG Car. Eccentric orbits are expected to rapidly circularise due to tidal interaction \citep{Zahn1977TidalStars}; however, both \cite{Soker2000EccentricStars} and \cite{BonacicMarinovic2008OrbitalStar} showed that high eccentricities can be sustained by enhanced mass-loss focused around periastron passages in tidally interacting systems where the mass-losing star is an evolved giant. Similarly, \cite{Sepinsky2007InteractingTransfer, Sepinsky2009INTERACTINGTRANSFER, Sepinsky2010InteractingSelf-accretion} have also shown that mass transfer focused around periastron can reverse the circularisation of binary orbits and pump up the eccentricity. Interaction of the binary with the system's circumbinary disk may also aid in sustaining the eccentricity of the binary orbit (e.g. \citealt{Dermine2013, Antoniadis2014}).} \\

\subsection{Binary evolutionary scenario for GG Car}
\label{sec:binary_evolution}
\cite{Kraus2013} pointed to GG Car's circumbinary $^{12}$C/$^{13}$C ratio, orbital eccentricity, the observation of \cite{Thackeray1950SomeNebulosity} of a lack of nebulosity, and the primary's theoretical age assuming a single star's evolutionary history as evidence that the primary evolved as a single star, and was a classical Be star previously in its main-sequence lifetime.  \\

However, recent studies have suggested that Be stars are overwhelmingly formed via binary interactions (e.g. \citealt{Shao2014ONINTERACTION, Klement2019PrevalenceBinaries, El-Badry2020ALB-1}). Furthermore, \cite{Bodensteiner2020InvestigatingStars} recently found a dearth of main sequence companions to classical Be stars, suggesting that mass transfer from the current secondaries and the associated spin-up of the current primaries dominates the Be star formation channel. Such interaction inverts the mass ratio of the binary, with the Be star being the initially less massive component. Therefore, assuming a Be progenitor of the primary of GG Car, these studies imply that mass is likely to have needed to have been transferred from the current secondary to the primary in the system's evolutionary history. \\

In this scenario, the $^{13}$C could have been formed in the now-secondary component during its evolution, allowing its surface to be enriched by the isotope; this then allows for the $^{13}$C to enrich the circumbinary disk, as non-conservative Roche lobe overflow provides a mechanism for the $^{13}$C enriched material to transcend from the secondary's surface to the circumbinary disk. Such non-conservative Roche lobe overflow is widely theorised to give rise to the circumbinary disks of \sgBeshorthand s (e.g. \citealt{Langer1998BeStatus}). \cite{Farrell2019ImpactSupergiants} notes that the high mass transfer rates during this interaction can both lead to dense circumstellar media that obscure the central stars, such as is observed in \sgBeshorthand s, and allow the spin up of the mass gainer. The spin up caused by this mass transfer may then allow decretion about the equator of the primary, thereby allowing the system to display the B[\,e\,] phenomenon described by \cite{Zickgraf1985}.\\

Such a binary evolutionary scenario leaves the originally more massive binary component less massive than its companion (e.g. \citealt{Wellstein2001FormationBinaries}). This evolutionary sequence has been well established in the study of high-mass X-ray binaries (HMXRBs; \citealt{vandenHeuvel1976LateSystems, vandenHeuvel2018High-MassObjects}) where it is expected the original primary would be a stripped He star before undergoing supernova to become a compact object. In this scenario, the now-primary would have been thrown out of thermal equilibrium by the mass transfer, and would have had a higher luminosity than expected for a single star of its mass; however, it would regain thermal equilibrium on its thermal timescale, $\tau_{\rm therm} = GM^2/(RL)\sim 4000$\,years. It is therefore unlikely that we are observing the system in this short-lived phase. \\

\textcolor{black}{In the previous section, we suggest that ongoing mass transfer focused at periastron may account for the high eccentricity seen in GG Car. However, HMXRBs have been observed to have high eccentricities \citep{Raguzova2005BeX-rayCandidates, Walter2015High-massWay}. Supernova kicks are suggested to account for the significant eccentricities in such systems \citep{Verbunt1995FormationBinaries.}. Such a scenario is not out of the realms of possibility for GG Car, if the current secondary were a supernova remnant. The mass of the secondary in the GG Car system is far larger than the maximum mass of a neutron star \citep{Rezzolla2018UsingStars}, so if the secondary were a supernova remnant it would need to be a black hole. The lack of X-rays observed in the system argues against this hypothesis, though it is possible that any X-rays produced by accretion would be obscured by GG Car's thick circumstellar environment and wind \citep{Koljonen2020The1915+105}.}\\

We cannot currently say more definitely about the exact evolutionary history of GG Car, since this would require the species of the secondary component in the system to be determined. This determination is precluded by the thick circumstellar environment, \textcolor{black}{but a search in the UV detecting the presence of a stripped He star would prove that mass transfer and mass inversion had occurred \citep{Gotberg2018SpectralStars}}. Overall, the evidence presented of ongoing of mass transfer from the primary to the secondary, the recent studies on the potential binary origin of the Be phenomenon, and studies that orbital eccentricity can survive binary interaction, demonstrate that the conclusion that the stars in GG Car evolved as single stars may not be accurate, and that the role of binarity should not be neglected when determining the evolution of systems such as GG Car.  \\

We therefore argue that the evolution of GG Car should be reevaluated in a binary context, in order to further understand the origin of its \sgBeshorthand\ phenomenon. \\

\section{Conclusions}
\label{sec:conclusions}
We have presented photometric and spectroscopic data of the \sgBeshorthand \ binary GG Car and studied its variability through the $\sim$31-day binary orbit of the system. \\

We have inferred a distance to GG Car of $d = 3.4^{+0.7}_{-0.5}$\,kpc using its parallax measured by \textit{Gaia} in DR2. We have determined the luminosity of the primary component, $L_{\rm pr} = 1.8^{+1.0}_{-0.7}\times10^5\,L_\odot$, and the radius of the primary, $R_{\rm pr} = 27^{+9}_{-7}\,R_\odot$, assuming the \cite{Marchiano2012} effective temperature $T_{\rm eff} = 23\,000 \pm 2000$\,K. Using the rotating and non-rotating stellar evolution models of \cite{Ekstrom2012GridsRotation}, we estimate the mass of the primary as $M_{\rm pr} = 24\pm4\,M_\odot$. \\ 

We have shown that H-alpha, He\,I, Si\,II and Fe\,II emission lines display radial velocity (RV) variability at the orbital period, and we have shown that the amplitudes of the emission lines' RV variations are correlated with the energy of the upper level of the line transition. This implies that the variations are correlated with the temperatures of the line forming regions. This is consistent with these emission lines being formed in the wind of the \sgBeshorthand\ primary. The wind traces the Keplerian motion of the primary, however lower energy lines form at larger radii in the wind on average, once it has been able to cool, than the higher energy lines, which form closer to the star where it is hotter. Hence there is a delay in emission of the different species according to the energies of the transitions in question. This effect thereby causes a smearing of the observed RV curves, with the lower energy lines smeared more than the higher energy lines. \\

\textcolor{black}{We have simulated the atmosphere of the \sgBeshorthand\ primary component of the binary GG Car using \texttt{CMFGEN} in order to calculate the emissivity of its H-alpha, He\,I, and Si\,II lines as a function of wind-flow time. We found that, in the simulation, these lines are formed in the wind of the primary, and that He\,I lines form closest into the stellar surface, on average, followed by H-alpha and then Si\,II lines which also form over a more extended region. This supports our assertion these lines form in the primary's wind, after finding the He\,I emission has the largest RV variations in GG Car, followed by H-alpha and Si\,II.} \\

\textcolor{black}{Using the simulated emissivity of these lines as a function of wind-flow time after passing the sonic point, we have been able to accurately measure the Keplerian motion of the primary component of the binary for the first time. We find that the system's orbit is considerably more eccentric ($e=0.50\pm0.03$) than previously thought. We define a new orbital ephemeris of the binary, given in Equation \ref{eq:ephemeris}. The newly determined orbital solution, convolved with the calculated line-formation kernels, is able to accurately reproduce the observed RV variations of the system's emission lines. Using limits of the circumbinary disk inclination of $60\pm20^\circ$ \citep{BorgesFernandes2010THEEYES, Kraus2013} and assuming the disk is roughly coplanar with the binary, we can constrain the mass of the secondary to $7.2^{+3.0}_{-1.3}\,M_\odot$ and the semi-major axis of the orbit to $a=0.61\pm0.03$\,AU. Therefore the mass ratio of the binary $q=M_{\rm 1}/M_{\rm 2} = 3.3^{+0.6}_{-1.0}$. Table \ref{tab:all_parameters} summarises all of the stellar and binary parameters of GG Car which have been determined in this study.}\\

\begin{table}
\centering          
\begin{tabular}{ l  l}
\hline
\hline \\
\textbf{Primary stellar parameters} & \\
$d$ & $3.4^{+0.7}_{-0.5}$\,kpc\\
$T_{\rm eff}$ & $23\,000 \pm 2000$\,K \\
$L_{\rm pr}$ & $1.8^{+1.0}_{-0.7}\times10^5\,L_\odot$ \\
$R_{\rm pr}$ & $27^{+9}_{-7}\,R_\odot$ \\
$M_{\rm pr}$  & $24\pm4\,M_\odot$ \\ 

\\
\textbf{Secondary parameters} & \\
$M_{\rm sec}$ & $7.2^{+3.0}_{-1.3}\,M_\odot$ \\

\\
\textbf{Orbital solution} & \\
$K$	& $48.57^{+2.04}_{-1.87}$\,\kms\\
$\omega$	& $339.87^{+3.10}_{-3.06}$\,$^\circ$	\\
$e$	& $0.50^{+0.03}_{-0.03}$	\\
$v_0$	& $-0.72^{+0.36}_{-0.36}$\,\kms	\\
$M_0$	& $202.35^{+15.16}_{-15.01}$\,$^\circ$ \\
$P$ & $31.01^{+0.01}_{-0.01}$\,days	\\
$T_{\rm peri}$ & $\,2452069.36\pm1.30$\,JD \\
$a$ & $0.61\pm0.03$\,AU\\
\end{tabular}
\caption{\textcolor{black}{Summary of the stellar and orbital parameters of GG Car determined in this study. \textit{Gaia} distance, $d$, and stellar parameters of the primary in GG Car, where $M_{\rm pr}$ is the mass of the primary, $T_{\rm eff}$ is the effective temperature of the primary, $L_{\rm pr}$ is the luminosity of the primary, and $R_{\rm pr}$ is the radius of the primary. $T_{\rm eff}$ is taken from \protect\cite{Marchiano2012}. $M_{\rm sec}$ is the mass of the secondary. $K$ is the orbit amplitude, $\omega$ is the argument of periastron, $e$ is orbital eccentricity, $v_0$ is the velocity offset without physical significance, $M_0$ is the phase offset of the reference time $T_0 = {\rm JD}\,2452051.93$, $P$ is the orbital period, $T_{\rm peri}$ is the epoch of periastron, and $a$ is the semi-major axis of the binary orbit.}} 
\label{tab:all_parameters}      
\end{table}

We have shown that the V-band photometry varies smoothly with one maximum and one minimum per orbital period with stochastic variations in amplitude superimposed upon these, such that the profiles of the lightcurve in successive orbits are non-identical. With the new orbital solution of the binary, we find that the phases of photometric maxima in GG Car occur just before periastron, and photometric minima occurring at apastron. Having discounted other sources of variability at the orbital period, we find that the shape of the V-band photometric variations of the system along the binary orbit can be reproduced using a model of enhanced mass transfer between the binary components at periastron and accretion of the primary's equatorial wind by the secondary. We therefore argue that the accretion of the primary's wind by the secondary is the cause of the photometric variations at the orbital period in GG Car. The stochastic nature of the shape of the V-band brightness variations can then be explained by variability in the wind of the \sgBeshorthand\ primary. \textcolor{black}{Mass-loss and mass-transfer from the primary focused at periastron would help explain the large eccentricity of the binary's orbit, which may otherwise be expected to have been circularised by tidal interaction.}\\

\textcolor{black}{We go on to show that the assumptions used to argue a single star evolutionary scenario for the primary in GG Car are flawed, and suggest that the evolution of the system ought to be reassessed in a binary paradigm.}\\

\section*{Acknowledgements}
We thank Philipp Podsiadlowski and John Papaloizou for their useful discussions. We are grateful to the anonymous referee whose constructive comments improved the quality of the manuscript. We thank John Hillier for useful discussions and guidance in using \texttt{CMFGEN}. AJDP thanks the Science \& Technology Facilities Council (STFC) for their support in the form of a DPhil scholarship. Part of this work was based on data from the OMC Archive at CAB (INTA-CSIC), pre-processed by ISDC. A great many organisations and individuals have contributed to the success of the Global Jet Watch observatories and these are listed on {\tt www.GlobalJetWatch.net} but we particularly thank the University of Oxford and the Australian Astronomical Observatory. This research has made use of NASA's Astrophysics Data System. This research has made use of the SIMBAD database, operated at CDS, Strasbourg, France.  This work made use of data supplied by the UK Swift Science Data Centre at the University of Leicester. This work has made use of data from the European Space Agency (ESA) mission
{\it Gaia} (\url{https://www.cosmos.esa.int/gaia}), processed by the {\it Gaia}
Data Processing and Analysis Consortium (DPAC,
\url{https://www.cosmos.esa.int/web/gaia/dpac/consortium}). Funding for the DPAC
has been provided by national institutions, in particular the institutions
participating in the {\it Gaia} Multilateral Agreement.

\section*{Data availability}
SWIFT 1SXPS X-ray data can be accessed from \url{http://www.swift.ac.uk/1SXPS/}.\\
\textit{Gaia} parallax data available from \url{https://gea.esac.esa.int/archive/}.\\
ASAS V-band photometric data available from \url{http://www.astrouw.edu.pl/cgi-asas/asas_cgi_get_data?105559-6023.5,asas3}.\\
ASAS-SN V-band photometric data available from \url{https://asas-sn.osu.edu/}.\\
OMC V-band photometric data available from \url{https://sdc.cab.inta-csic.es/omc/secure/form_busqueda.jsp}.\\
FEROS spectroscopic data available from ESO at \url{http://archive.eso.org/scienceportal/home} and \url{http://dc.zah.uni-heidelberg.de/feros/q/web/form}.\\
The fits to spectroscopic Global Jet Watch data underlying this article will be shared on reasonable request to the corresponding author.

\bibliography{references_new}{}
\bibliographystyle{mnras}

\appendix
\section{Spectroscopic observations}
\begin{figure} 
  \centering
    \includegraphics[width=0.5\textwidth]{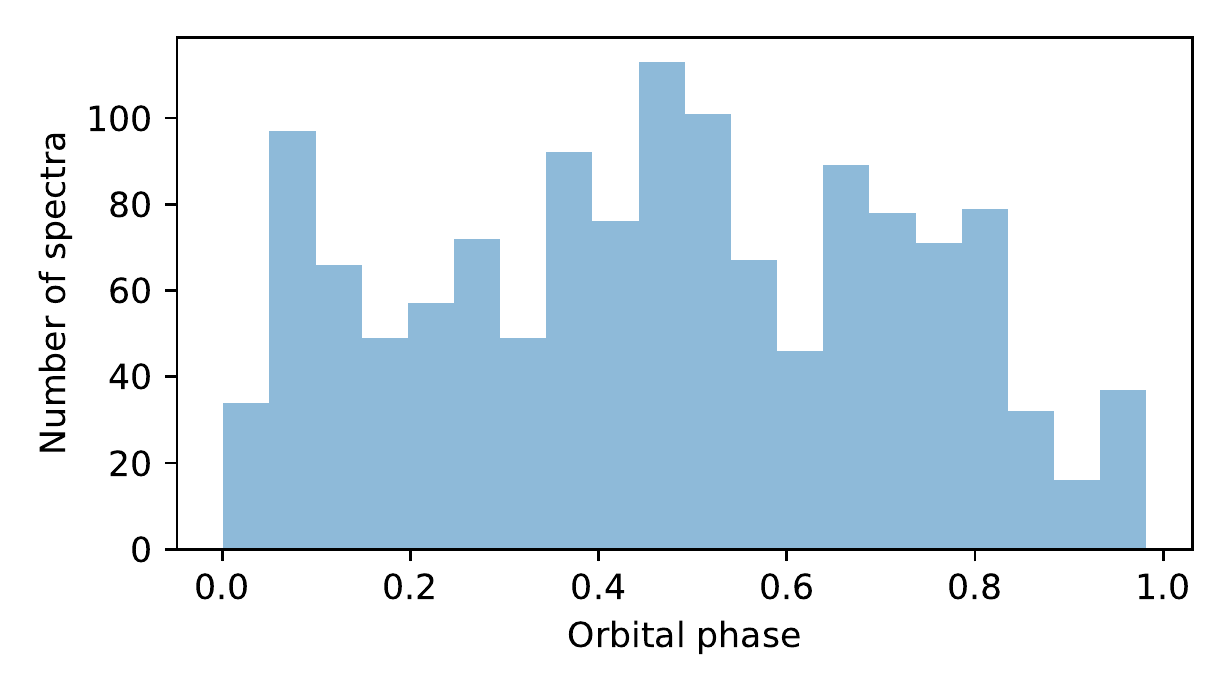}
    \caption{The number of GJW spectra, as a function of the orbital phase of GG Car. The phases are calculated using Equation \ref{eq:ephemeris}.}
    \label{fig:n_spectra_by_phase}
\end{figure}

Figure \ref{fig:n_spectra_by_phase} displays the distribution of GJW spectra as a function of orbital phase of the binary in GG Car. \\

\begin{figure} 
  \centering
    \includegraphics[width=0.5\textwidth]{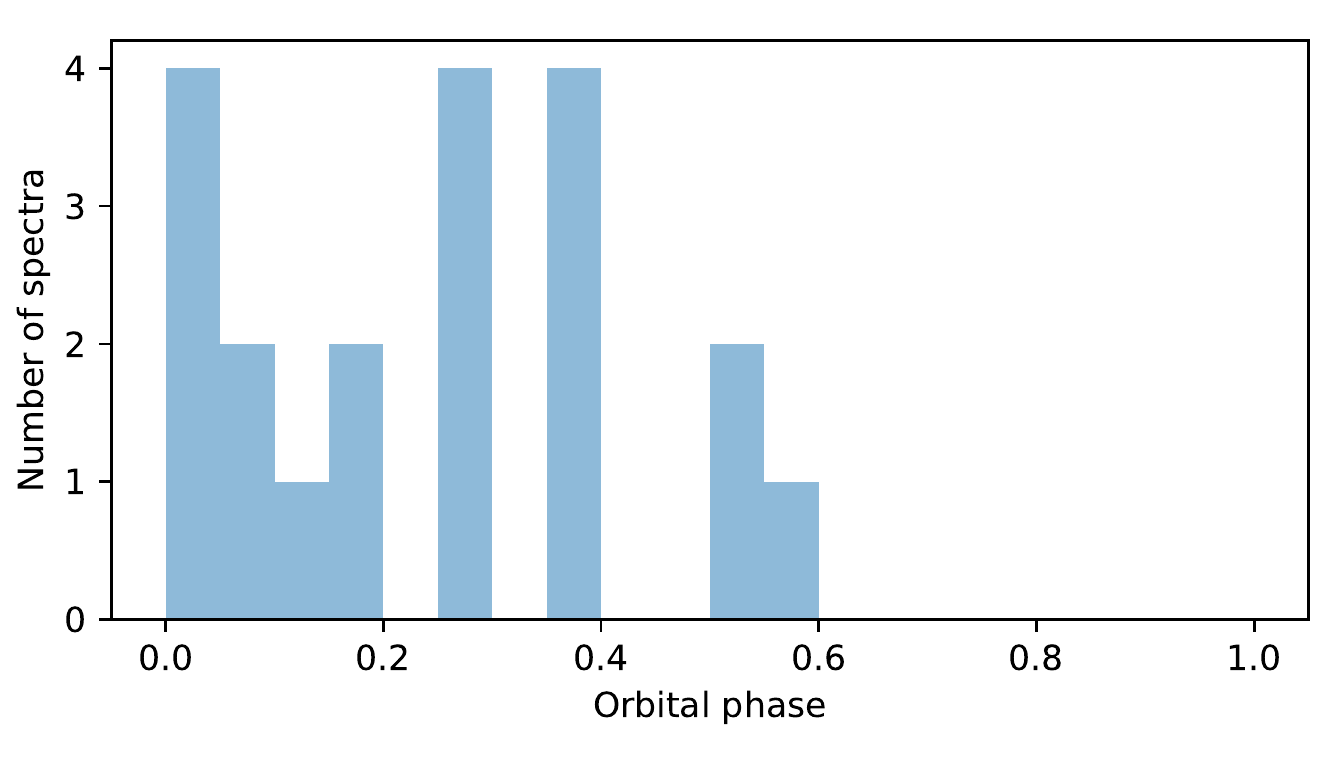}
    \caption{Same as Figure \ref{fig:n_spectra_by_phase}, except for the publicly available FEROS spectra, listed in Table \ref{tab:feros_observations}.}
    \label{fig:feros_phase_coverage}
\end{figure}

Table \ref{tab:feros_observations} lists the FEROS spectra of GG Car which are publicly available, and Figure \ref{fig:feros_phase_coverage} shows how these observations are distributed along the orbital phase of GG Car.\\

\begin{table} 
\centering          
\begin{tabular}{ l c}
\hline \hline
\\
Observation date & Julian Date \\
\hline
\\
1998-12-07	&	2451154.813	\\
1998-12-08	&	2451155.857	\\
1998-12-09	&	2451156.864	\\
1998-12-24	&	2451171.759 \\
1998-12-25	&	2451172.809	\\
1998-12-26	&	2451173.728	\\
1998-12-27	&	2451174.725	\\
1998-12-28	&	2451175.729	\\
2015-05-13	&	2457155.651	\\
2015-05-13	&	2457155.653	\\
2015-05-13	&	2457155.655	\\
2015-05-13	&	2457155.657	\\
2015-11-23	&	2457349.859	\\
2015-11-23	&	2457349.861	\\
2015-11-23	&	2457349.863	\\
2015-11-23	&	2457349.865	\\
2015-11-26	&	2457352.849	\\
2015-11-26	&	2457352.851	\\
2015-11-26	&	2457352.853	\\
2015-11-26	&	2457352.855	\\
  \hline
\end{tabular}
\caption{The FEROS observations of GG Car.} 
\label{tab:feros_observations}      
\end{table}

\section{Orbital solution fits}
Table \ref{tab:orbital_solutions} displays the results of the orbital solution fits for the RV variations at the orbital period for the individual emission lines that were studied, along with the wavelengths, energies of the lower atomic state ($E_i$) and higher atomic state ($E_k$) associated with the line transitions. The wavelengths and energies were accessed from the NIST Atomic Spectra Database.\\

\begin{table*} 
\centering          
\begin{tabular}{ l l l l l l l l l}
\hline \hline
\\
Species and wavelength (\AA)	&	K (\kms)	&	$\omega$ ($^\circ$)	&	e	&	$v_0$ (\kms)		&	$M_0$ ($^\circ$) &	jitter (\kms) & $E_i$ (eV) & $E_k$ (eV)	\\
\hline \\
Fe$\,$II$\,$6317.3871	&	$9.98^{+0.89}_{-0.79}$	&	$311^{+3.9}_{-5.1}$	&	$0.271^{+0.068}_{-0.074}$	&	$-5.1^{+0.48}_{-0.44}$	&	$36.7^{+0.86}_{-2.1}$	&	$9.73^{+0.34}_{-0.33}$	&  6.2224968  &  8.1845409\\
Fe$\,$II$\,$6383.7302	&	$10.4^{+0.62}_{-0.62}$	&	$296^{+11}_{-11}$	&	$0.286^{+0.059}_{-0.061}$	&	$-3.6^{+0.4}_{-0.39}$	&	$22^{+9.9}_{-9.4}$	&	$9.69^{+0.27}_{-0.26}$	& 5.55260574  &  7.4942594\\
Fe$\,$II$\,$6456.3796	&	$7.7^{+0.57}_{-0.57}$	&	$335^{+27}_{-20}$	&	$0.214^{+0.078}_{-0.074}$	&	$-23.8^{+0.45}_{-0.38}$	&	$47.4^{+26}_{-18}$	&	$8.83^{+0.3}_{-0.24}$	&  3.90341902  &  5.8232247 \\
Fe$\,$II$\,$6491.663	&	$12.3^{+1.3}_{-1.2}$	&	$260^{+12}_{-11}$	&	$0.457^{+0.08}_{-0.078}$	&	$0.666^{+0.62}_{-0.61}$	&	$359^{+9.2}_{-7.4}$	&	$15.5^{+0.44}_{-0.43}$	&  10.9088323  &  12.8182036\\
Fe$\,$II$\,$7513.1762	&	$8.12^{+1}_{-0.93}$	&	$276^{+97}_{-79}$	&	$0.0975^{+0.17}_{-0.078}$	&	$0.968^{+0.64}_{-0.74}$	&	$18.9^{+1.5e+02}_{-38}$	&	$13.5^{+0.53}_{-0.43}$	&  9.6536142  &  11.3033834\\
Fe$\,$II$\,$7711.4386	&	$6.78^{+0.79}_{-0.66}$	&	$258^{+15}_{-15}$	&	$0.392^{+0.077}_{-0.078}$	&	$-24.8^{+0.45}_{-0.35}$	&	$350^{+11}_{-13}$	&	$7.88^{+0.41}_{-0.26}$	&  5.51071386  &  7.1180674\\
H alpha\,-\,6562.819	&	$18.9^{+2}_{-1.9}$	&	$262^{+19}_{-21}$	&	$0.257^{+0.096}_{-0.12}$	&	$20.6^{+1.5}_{-1.2}$	&	$345^{+19}_{-23}$	&	$17.9^{+1.1}_{-1}$	&  10.1988357  &  12.0875051\\
He$\,$I$\,$5875.5987	&	$41^{+1.9}_{-1.8}$	&	$12.3^{+15}_{-13}$	&	$0.247^{+0.045}_{-0.05}$	&	$62.3^{+1.4}_{-1.2}$	&	$73.7^{+14}_{-12}$	&	$28.9^{+0.95}_{-0.84}$	&  20.96408703  &  23.07365663\\
He$\,$I$\,$6678.1517	&	$46^{+2.8}_{-2.5}$	&	$313^{+7.2}_{-6}$	&	$0.445^{+0.048}_{-0.057}$	&	$77.3^{+1.5}_{-1.6}$	&	$27.5^{+4.8}_{-4.2}$	&	$29.3^{+0.9}_{-0.99}$	&  21.21802284  &  23.07407493\\
He$\,$I$\,$7065.17714	&	$32.8^{+4.2}_{-3.9}$	&	$336^{+34}_{-62}$	&	$0.179^{+0.16}_{-0.13}$	&	$86.8^{+3.1}_{-3.1}$	&	$29.5^{+42}_{-37}$	&	$50.1^{+2.1}_{-2}$ &  20.96408703  &  22.71846655	\\
Si$\,$II$\,$5957.56	&	$16.4^{+2.8}_{-2.6}$	&	$298^{+17}_{-18}$	&	$0.431^{+0.1}_{-0.15}$	&	$-86.2^{+1}_{-1.1}$	&	$37.2^{+12}_{-12}$	&	$23.6^{+0.84}_{-0.8}$	&  10.066443  &  12.146991\\
Si$\,$II$\,$5978.93	&	$24.9^{+1.6}_{-1.6}$	&	$302^{+7.2}_{-7.2}$	&	$0.501^{+0.047}_{-0.049}$	&	$-48.8^{+0.84}_{-0.85}$	&	$27.1^{+4.6}_{-4.2}$	&	$20.9^{+0.64}_{-0.6}$ &  10.07388  &  12.146991	\\
Si$\,$II$\,$6347.11	&	$25^{+0.98}_{-0.99}$	&	$333^{+8}_{-7.4}$	&	$0.311^{+0.034}_{-0.032}$	&	$-41.7^{+0.57}_{-0.62}$	&	$49.4^{+6.8}_{-6.4}$	&	$13.8^{+0.46}_{-0.4}$	&  8.121023  &  10.07388\\
Si$\,$II$\,$6371.37	&	$24^{+1.6}_{-1.5}$	&	$287^{+9.3}_{-8.3}$	&	$0.462^{+0.05}_{-0.045}$	&	$-62.8^{+0.83}_{-1}$	&	$22^{+6.5}_{-5.4}$	&	$17.2^{+0.76}_{-0.8}$ &  8.121023  &  10.066443	\\
\end{tabular}
\caption{The line identifications and orbital solutions fitted to the 31-day RV variations for each emission line used in this study, along with the energy of the lower atomic state, $E_i$, and energy of the upper atomic state, $E_k$, of the lines. Wavelengths and energies were accessed from the NIST Atomic Spectra Database.} 
\label{tab:orbital_solutions}      
\end{table*}

\section{Fe\,II disk profiles}
\begin{figure} 
  \centering
    \includegraphics[width=0.5\textwidth]{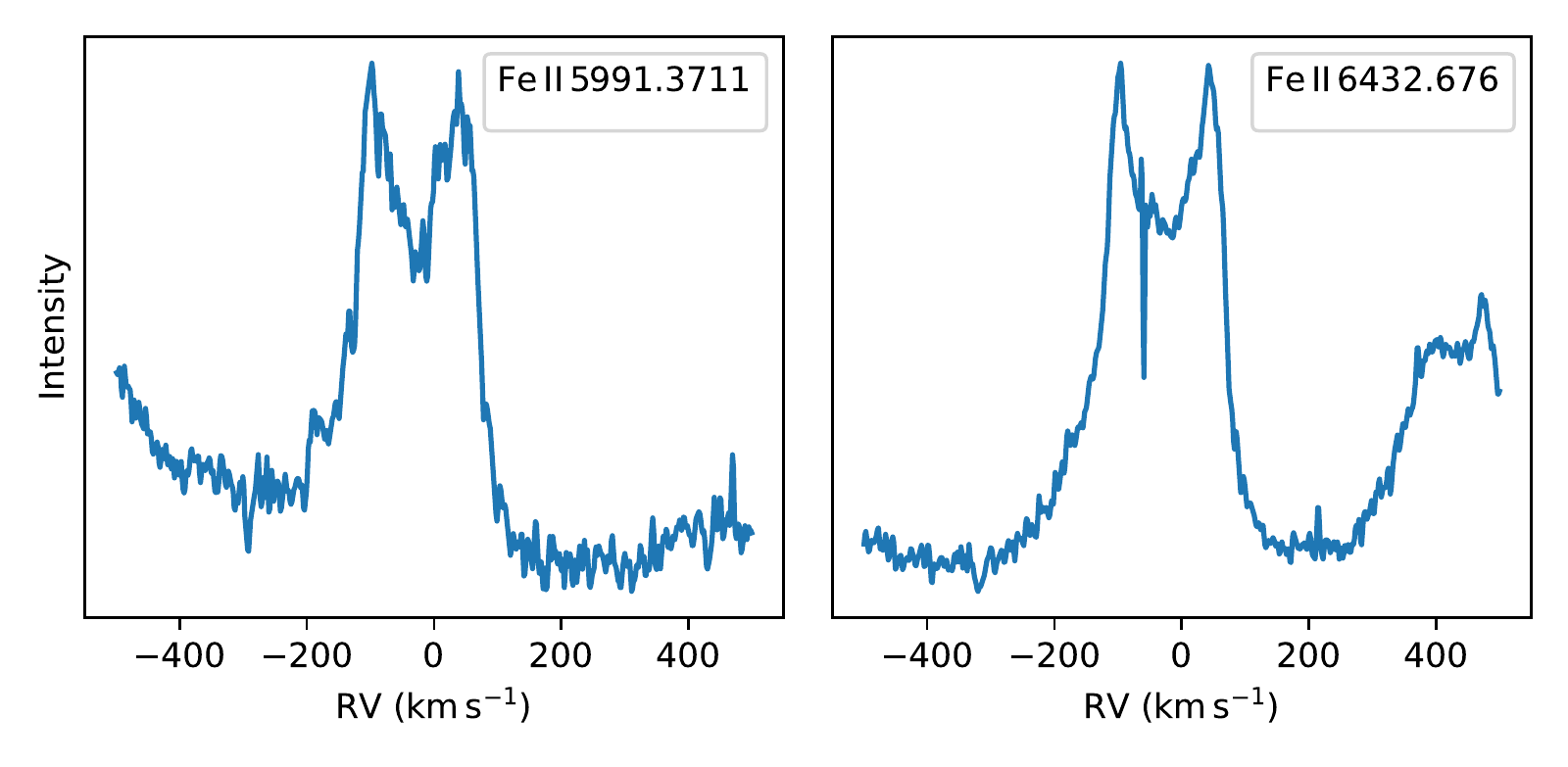}
    \caption{Same as Figure \ref{fig:feros_si_lines}, except for the Fe\,II lines which do not exhibit variability at the orbital period.}
    \label{fig:feros_fe_disk_lines}
\end{figure}

Figure \ref{fig:feros_fe_disk_lines} displays the FEROS observations of the Fe\,II lines which do not display RV variability at the orbital period. Their line profiles clearly resemble double-peaked disk or ring profiles with a peak-to-peak splitting of $\sim$160\,km\,s$^{-1}$ and a central RV of $\sim -20$\,\kms. \cite{Kraus2013} and \cite{Maravelias2018} discovered circumbinary material orbiting from $\sim$26 to $\sim$80\,$\rm km\,s^{-1}$, projected along the line of sight, placing these Fe\,II lines in those regions. Therefore, due to their disk profiles, splitting, and lack of RV variability, we ascribe these lines as forming in the circumbinary disk of the system. These two lines are both semi-forbidden and have low values of $E_k$ (5.2\,eV and 4.8\,eV for Fe\,II 5991 and 6432, respectively) which may explain why they are only formed in the system's circumbinary disk. These circumbinary emission lines will be studied in more detail in an upcoming paper.\\

\section{Relationship of $v_0$ and jitter with $E_k$}
\label{sec:v0_v_ek}

\begin{figure} 
  \centering
    \includegraphics[width=0.5\textwidth]{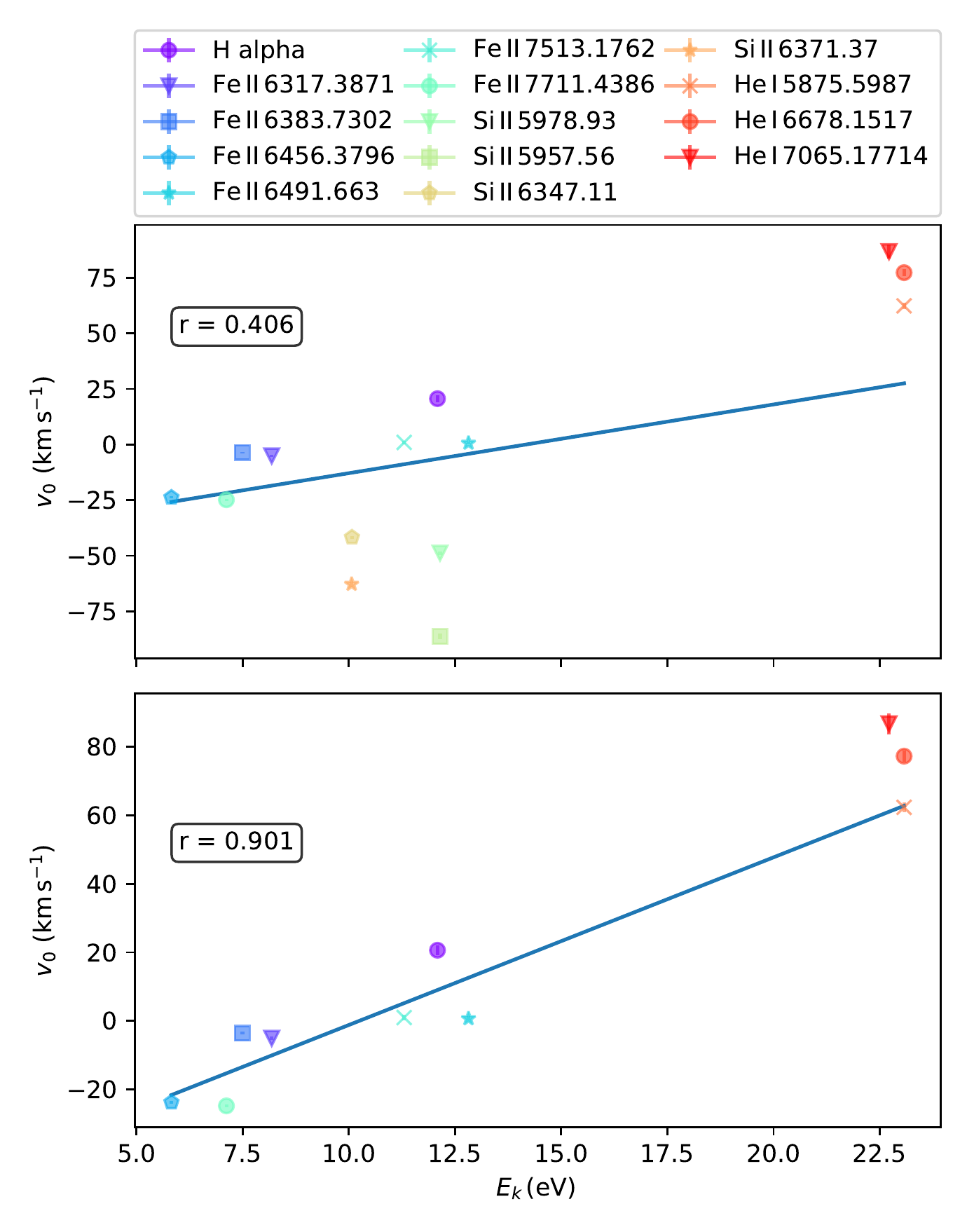}
    \caption{The relationship of $v_0$ with $E_k$ for all lines investigated in this study (top panel) and for all lines except Si\,II (bottom panel). The correlation parameter, $r$, is given in the panel for both of these cases.}
    \label{fig:v0_v_ek}
\end{figure}

\textcolor{black}{Figure \ref{fig:v0_v_ek} displays the relationship of systemic velocity, $v_0$, of the emission lines which exhibit RV variability versus $E_k$, the energy of the upper atomic state. The top panel shows the relationship for all lines, and the bottom panel shows the relationship for all lines except for the Si\,II species. In the top panel, there is no statistically significant relationship between the two parameters. In the bottom panel, once Si\,II has been excluded, there is a very strong correlation between the parameters with higher $E_k$ being correlated with higher redshifts. The minimal $v_0$s, of the Fe\,II\,6456 and 7711 lines, closely match the systemtic velocity of the system reported by \cite{Maravelias2018}.} \\

\textcolor{black}{Despite being an interesting result, we cannot currently say further whether the correlation between $v_0$ and $E_k$ is true and the Si\,II lines are outliers which are more blueshifted than they should be, or whether there is no real relationship at all between $v_0$ and $E_k$.} \\

\begin{figure} 
  \centering
    \includegraphics[width=0.5\textwidth]{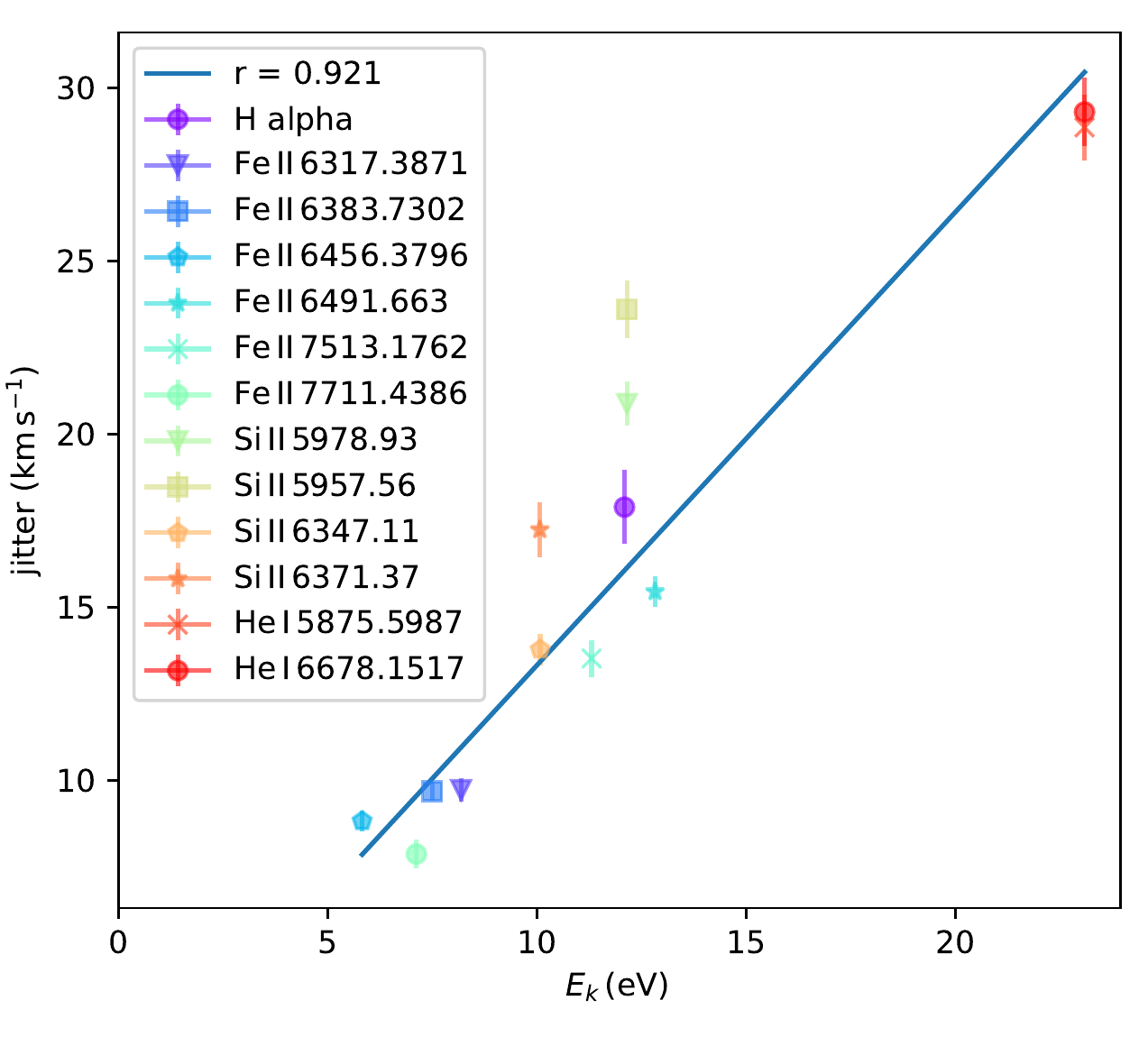}
    \caption{The relationship of jitter with $E_k$ for all lines investigated in this study, except He\,I 7065 which is excluded because its jitter is anomalously large due to the more complex fitting routine that must be used to extract the line's RVs. The correlation parameter, $r$, is given in the legend.}
    \label{fig:jitter_v_ek}
\end{figure}

\textcolor{black}{Figure \ref{fig:jitter_v_ek} displays the relationship of jitter against $E_k$. There is a very clear relationship between these two parameters, with higher $E_k$ leading to more jitter needed in the fitting routine. This indicates that higher $E_k$ is related to more scatter in the RV variations of the emission lines over the orbital period. We exclude He\,I 7065 from this analysis since the jitter for this line He\,I 7065 is anomalously larger than the other He\,I lines. This is due to the more complex fitting routine that is required to extract the RV of the He\,I 7065 emission; this introduces scatter in the RV curve due to fitting inaccuracies which inflates the measured jitter.}\\

\section{Determining $v_{\infty}$}
\label{sec:v_inf}
The terminal wind velocity, $v_{\infty}$, is used in the \texttt{CMFGEN} simulation of GG Car's atmosphere in order to determine the line-emissivity kernels. As \sgBeshorthand s have non-spherical winds, but instead host both a fast, polar wind and a slow equatorial wind, we use the one-dimensional code to simulate the equatorial wind since the lines we are studying are expected to form in that component.\\

\begin{figure}
  \centering
    \includegraphics[width=0.5\textwidth]{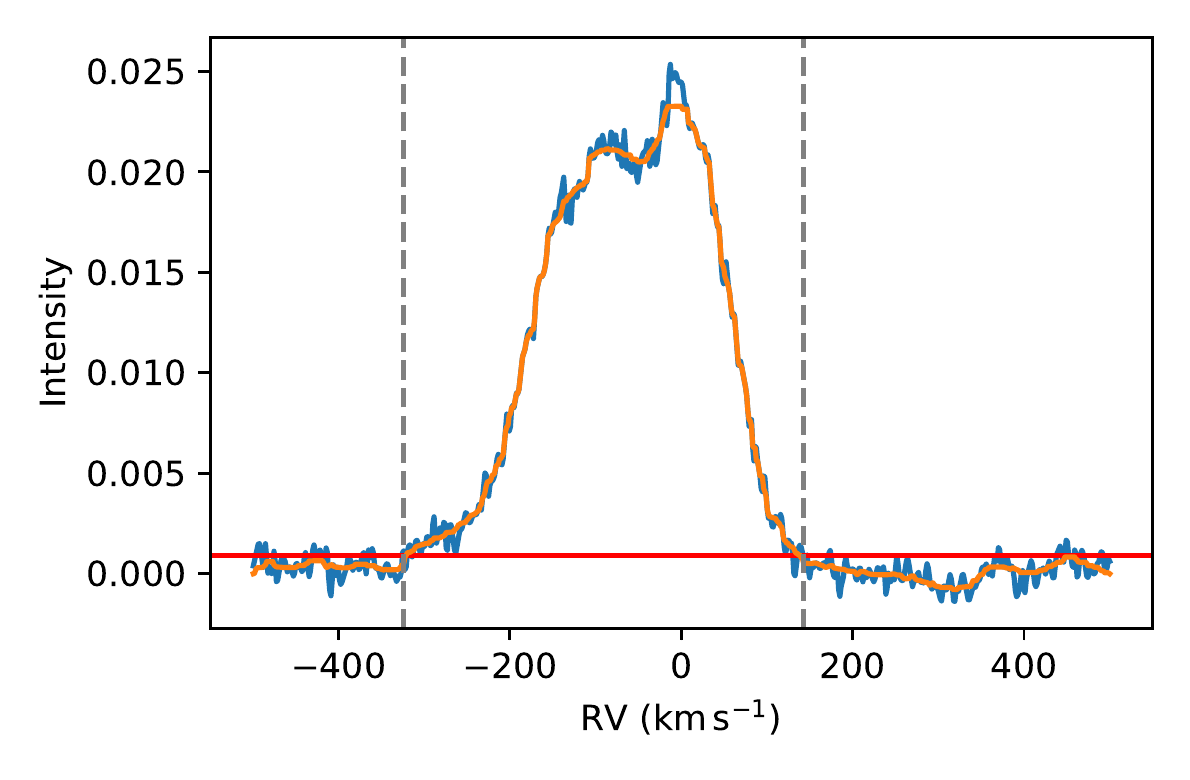}
    \caption{Determination of the HWZI of an Si\,II line in FEROS. The original spectrum is in blue, and the smoothed spectrum is in orange. A red line indicates where we define zero intensity, and dashed grey lines indicate where the smoothed spectrum drops below this level. The HWZI is half the distance between the two dashed lines.}
    \label{fig:fwzm}
\end{figure}

The first determination of $v_\infty$ comes from the half-width-at-zero-intensity (HWZI) of the Si\,II 6347 line in the FEROS spectra. This line is the cleanest of all the lines used in this study, having just one major component and no absorptions, unlike the He\,I and H-alpha lines.  Si\,II 6371, unfortunately, suffers some mild blending which slightly affects the width determination of the line. The HWZI for each FEROS spectrum was determined by taking a median-filter of the Si\,II line in order to get a smoothed line profile. The noise of the spectrum was then taken as the mean of the absolute differences between the spectrum and the smoothed spectrum. We then define the line as being at zero-intensity when the smoothed spectrum dips below the level of the continuum plus the noise. Figure \ref{fig:fwzm} displays an example of determining the HWZI of an Si\,II line. We calculate the HWZI of the Si\,II 6347 line in all of the available FEROS spectra. The width is variable across the FEROS dataset, indicating wind fluctuations, but we find a median HWZI of 264.67\,\kms\ with a standard deviation of 21\,\kms. This width is comparable to the widths of the He\,I and H\,I Balmer lines.\\

\begin{figure}
  \centering
    \includegraphics[width=0.5\textwidth]{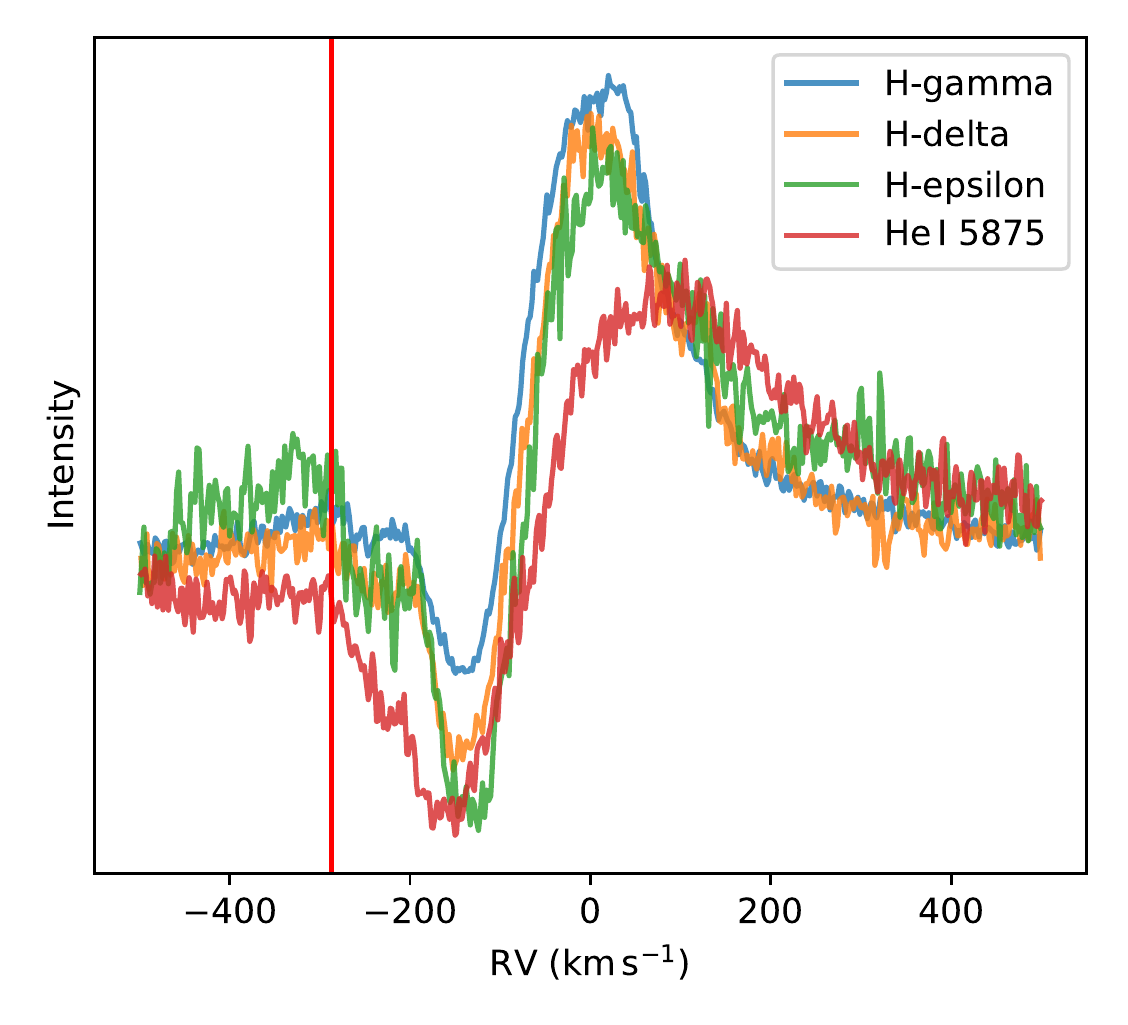}
    \caption{Balmer H\,I complexes and He\,I 5875 in an example FEROS spectrum of GG Car. The spectrum has been continuum subtracted, then each line normalised by the standard deviation of the spectrum in the velocity range. A vertical red line is placed at $-$287\,\kms .}
    \label{fig:h_he_absorption}
\end{figure}

The second determination of $v_\infty$ comes from the position of the bluemost edge of the middle absorption component in the Hydrogen Balmer series and He\,I lines in the FEROS spectra. Figure \ref{fig:h_he_absorption} displays three Balmer Hydrogen lines and the He\,I 5875 line in a FEROS spectrum of GG Car. Generally these lines show three distinct absorption components, with the less blueshifted two often blended. Their bluemost edges are around $\sim -$160, $\sim-$290 and $\sim-$550\,\kms. The absorption around $-160$\,\kms\ matches the H-alpha absorption visible in the GJW spectra, shown in Figure \ref{fig:halpha_trailed_spectra}, and shows little sign of RV variability. We focus on the component around $-$290\,\kms, since this is similar to the Si\,II HWZI. As with the Si\,II 6347 HWZI, the exact RV varies slightly over the FEROS observations. This will be due to both variations in the wind conditions and any residual motion of the wind when it was ejected from the \sgBeshorthand\ primary. Since the systemic velocity is $\sim -$22\,\kms, average speed of this wind component is $\sim -$265\,\kms.\\

Therefore, we set $v_\infty$ in the \texttt{CMFGEN} simulation to be 265\,\kms. Since we observe this to be the characteristic HWZI of the Si\,II 6347 line and due to the blue-most edge of the He\,I and H\,I Balmer absorptions having this same RV, this gives evidence that these lines form in the same wind component. This 265\,\kms\ is an estimate of the average $v_\infty$ of this wind component, even though we observe it to vary slightly over the FEROS dataset. \\

\section{Posterior distribution of orbital solution fit}
\begin{figure*}
  \centering
    \includegraphics[width=0.9\textwidth]{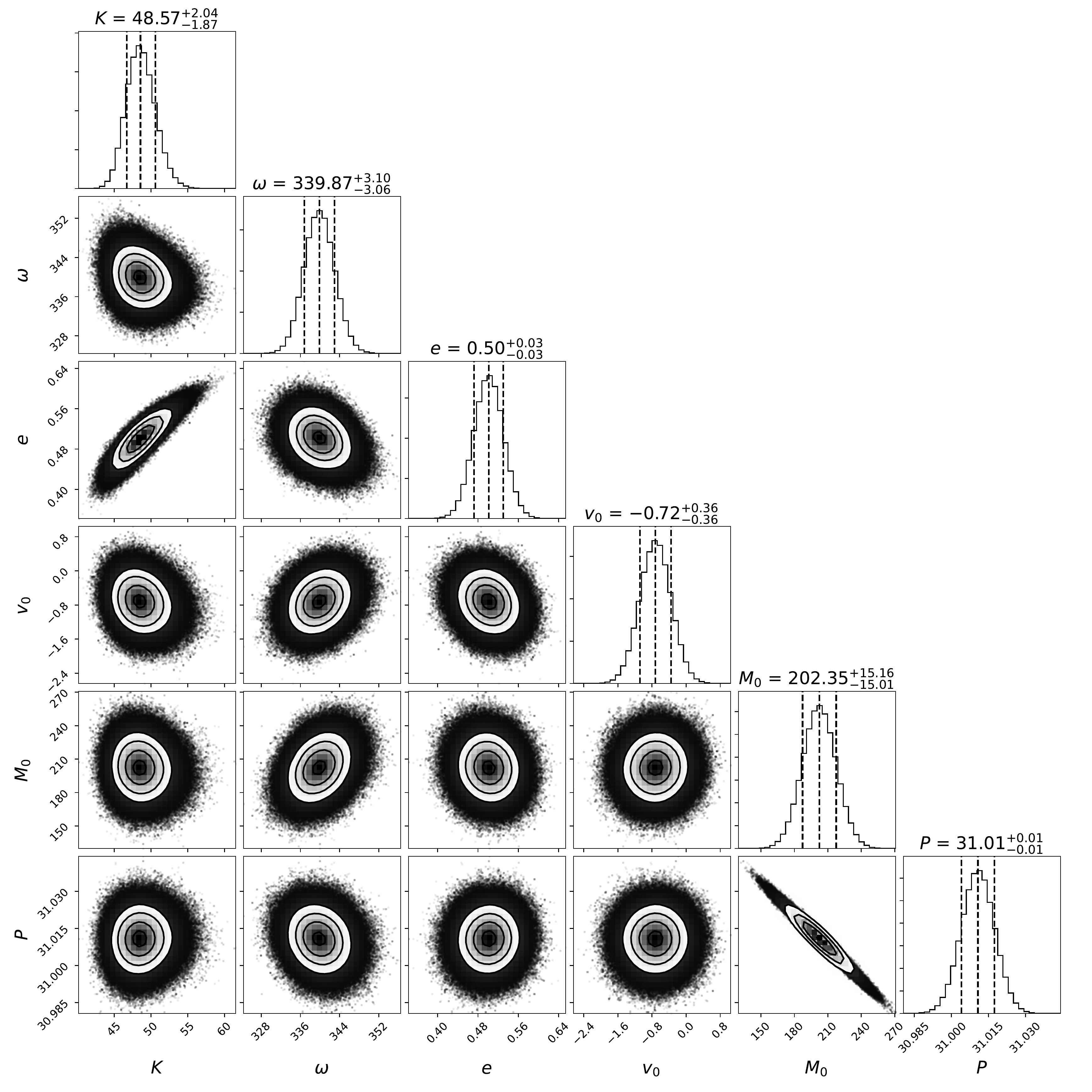}
    \caption{The posterior probability distribution for the parameters in the fits of the Keplerian orbital solution. Dashed vertical lines denote the 16th, 50th, and 84th percentiles. $K$ is the orbit amplitude, $\omega$ is the argument of periastron, $e$ is orbital eccentricity, $v_0$ is the velocity offset without physical significance, $M_0$ is the phase offset of the reference time $T_0 = {\rm JD}\,2452051.93$, and $P$ is the orbital period.}
    \label{fig:convolved_fitted_keplerian_corner}
\end{figure*}

Figure \ref{fig:convolved_fitted_keplerian_corner} displays the posterior distribution of the MCMC fit of the orbital solution, utilising the \texttt{CMFGEN} line emissivity kernels, described in Section \ref{sec:orbital_solution_results}.\\

\section{Accretion model}
\label{sec:accretion_model}
\textcolor{black}{In the two-dimensional wind-accretion model described in Section \ref{sec:enhanced_mass_transfer}, we calculate the position of the two binary components using Kepler's laws and the binary orbital parameters measured in Section \ref{sec:determining_orbital_solution}. The primary emits a radial wind from its equator with speed $v_{\rm wind}$ from the reference frame of the primary, which is modelled as a discrete ring being ejected at every timestep, $dt$. The centre of the ring travels with a bulk velocity equal to the velocity of the primary at ejection, $\bar{v}_{\rm bulk}$. This makes the center of the ``wind ring''}
\begin{equation}
    \bar{x}_{\rm ring} = \bar{x_0} + \bar{v}_{\rm bulk}\,t_{\rm ring},
\end{equation}

\noindent \textcolor{black}{where $\bar{x_0}$ is the location of the primary when the ring was ejected and $t_{\rm ring}$ is the life time of the ring since emission. The radius of the ring is}
\begin{equation}
    r_{\rm ring} = R_* + v_{\rm wind}\,t_{\rm ring},
\end{equation}

\noindent \textcolor{black}{where $R_*$, the stellar radius, is the radius at which the ring is ejected.} \\

\textcolor{black}{Matter within the ring is counted as accreted by the secondary when the expanding ring passes within the Roche lobe of the secondary. This occurs when}
\begin{equation}
    r_{\rm ring} > |\bar{x}_{\rm ring} - \bar{x}_{\rm comp}| - r_{\rm roche},
\end{equation}

\noindent \textcolor{black}{where $\bar{x}_{\rm comp}$ is the position of the companion at the timestep, and $r_{\rm roche}$ is the radius of the secondary's Roche lobe, calculated using Equation \ref{eq:roche} and the instantaneous separation of the components in their orbit. As we have assumed the wind is radial, at accretion the wind will be travelling directly away from $\bar{x}_{\rm ring}$ with a magnitude $v_{\rm wind}$; this velocity vector is then $\bar{v}_{\rm ring}$. The velocity difference between the wind and the secondary is}
\begin{equation}
    v_{\rm diff} = |\bar{v}_{\rm comp} - \bar{v}_{\rm ring}|,
\end{equation}

\noindent \textcolor{black}{where $\bar{v}_{\rm comp}$ is the velocity of the companion at accretion.}\\

\textcolor{black}{To determine the amount of mass accreted per ring, we build a first-order weighting from first principles to be applied when the radius of the ring passes the Roche radius of the secondary. Since we assume that the wind of the primary is concentrated in a thin disk in the orbital plane, we consider a two-dimensional problem of a mass of accretion radius $R_A$ travelling through a medium of mass per unit area $\rho$ at velocity $v$. The mass accretion rate, $\dot{M}$, will be equal to}
\begin{equation}
\label{eq:bondi_hoyle_2d}
    \dot{M} = 2\,R_A\,\rho\,v.
\end{equation}

\noindent \textcolor{black}{The accretion radius $R_A$ is the critical impact parameter at which a particle travelling with speed $v$ toward the accretor will be accreted. $R_A$ can be found by equating the gravitational energy of the particle with its kinetic energy, and is given by}
\begin{equation}
\label{eq:accretion_radius}
    R_A = \frac{2\,G\,M}{v^2}.
\end{equation}

\noindent \textcolor{black}{Assuming the wind is composed of many infinitesimally thin rings each of mass $m$, radius $r$, and thickness $dr$, the mass of the ring per unit area is}
\begin{equation}
    \label{eq:density_2d}
    \rho = \frac{m}{2\pi r \,dr}. 
\end{equation}

\noindent \textcolor{black}{Combining Equations \ref{eq:bondi_hoyle_2d}, \ref{eq:accretion_radius}, and \ref{eq:density_2d}, we find that the accretion rate can be described by the proportionality}
\begin{equation}
\label{eq:accretion_proportionality}
    \dot{M} \propto \frac{1}{rv}.
\end{equation}

\noindent \textcolor{black}{This gives the weighting which we quote in Equation \ref{eq:accretion_weighting}.} \\

\begin{figure}
  \centering
    \includegraphics[width=0.5\textwidth]{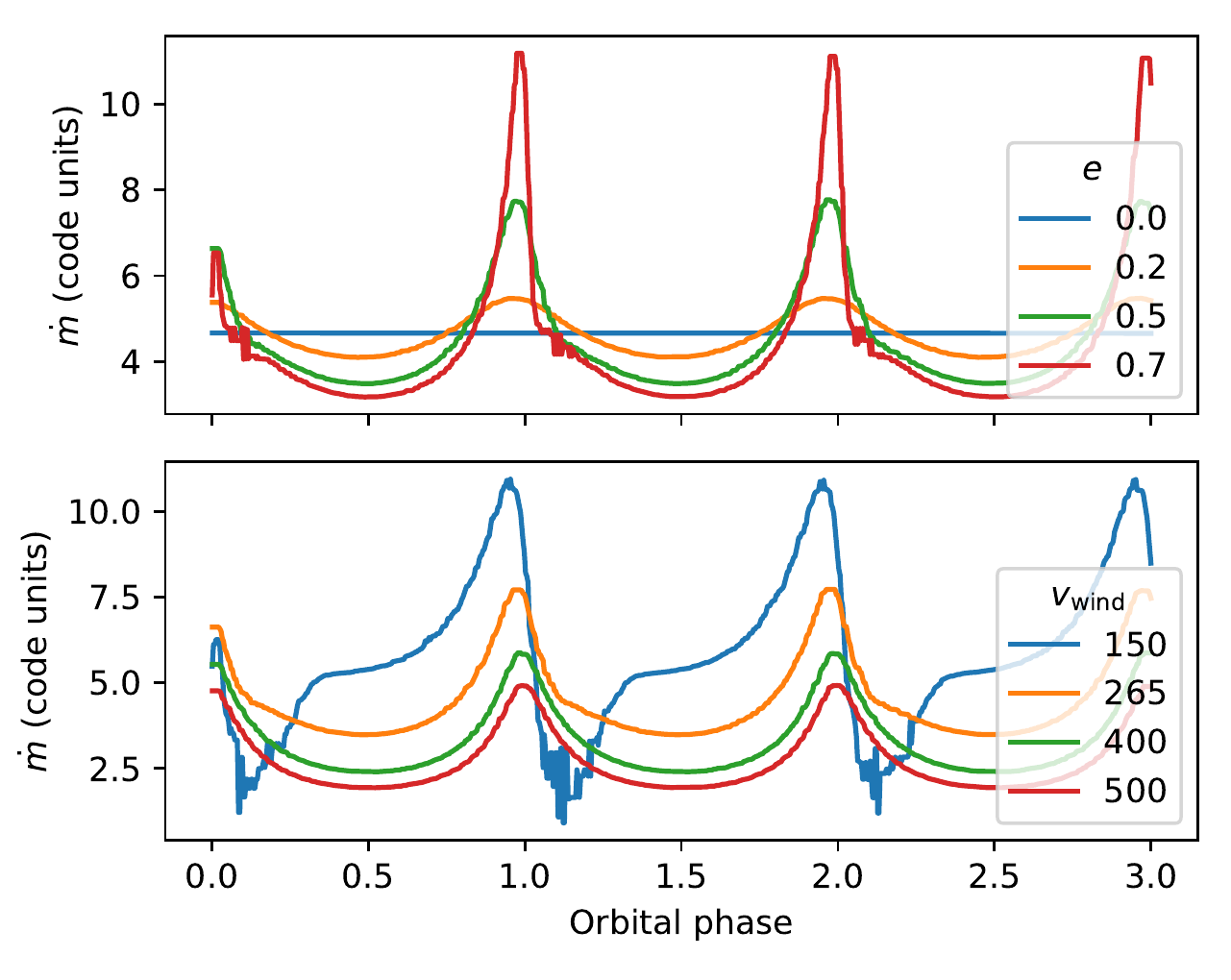}
    \caption{Variations in the modelled accretion rate, $\dot{m}$ over the orbital period after varying the eccentricity $e$ (top panel) and wind velocity $v_{\rm wind}$ (bottom panel). All other calculation parameters are kept constant as described in Section \ref{sec:enhanced_mass_transfer}.}
    \label{fig:accretion_simulation_comparison}
\end{figure}

\textcolor{black}{Figure \ref{fig:accretion_simulation_comparison} compares how $\dot{m}$, the modelled accretion rate described in Section \ref{sec:enhanced_mass_transfer}, varies by changing the input eccentricity, $e$, or wind velocity, $v_{\rm wind}$. All parameters, other than those indicated in the legend, are kept constant in these calculations compared to the calculation in Section \ref{sec:enhanced_mass_transfer}. Unsurprisingly, lower $e$ leads to a more uniform accretion rate through the orbital period, culminating in a constant $\dot{m}$ for $e = 0$. Higher eccentricities lead to more ``pulsed'' accretion profiles, with relatively more accretion occurring right at periastron.}\\

\textcolor{black}{On the other hand, lowering $v_{\rm inf}$ below the 265\,\kms\ assumed for the calculation in Section \ref{sec:enhanced_mass_transfer} leads to an irregular profile of the accretion rate, with the rate peaking before periastron and then falling dramatically after periastron. Increasing the wind speed above 265\,\kms\ does not significantly affect the accretion signature, except $\dot{m}$ becomes more symmetric around periastron with the peak rate occurring at periastron.}\\

\textcolor{black}{These show that the accretion model is sensitive to the chosen $e$ and $v_{\rm wind}$. The eccentricity of the model with $e = 0.5$ is the closest match to the V-band photometric variations observed, but any eccentricity between $\sim 0.2$\,--\,0.7 would be acceptable. Likewise, below the chosen $v_{\rm wind}$ of 265\,\kms\ the calculated $\dot{m}$ does not resemble the V-band photometric variations observed. However, the calculation is robust with higher $v_{\rm wind}$.} \\

\bsp	
\label{lastpage}
\end{document}